\documentstyle[12pt,epsf]{article}
\newcommand{\lsc}{\Lambda _\chi}
\newcommand{\cpt}{\chi PT}
\newcommand{\sss}{\scriptscriptstyle }
\newcommand{\ga}{g_{\sss A}}

\newcommand{\vsigone}{{\vec\sigma^{\sss 1}}}
\newcommand{\vsigtwo}{{\vec\sigma^{\sss 2}}}
\newcommand{\veps}{{\vec\epsilon}}
\newcommand{\vepsprime}{{\vec\epsilon\, '}}
\newcommand{\vkay}{{\vec k}}
\newcommand{\vkayprime}{{{\vec k}\, '}}
\newcommand{\vpee}{{\vec p}}
\newcommand{\vpeeprime}{{{\vec p}\, '}}
\newcommand{\vsigma}{{\vec\sigma}}

\newcommand{\barf}{\Upsilon}

\begin{document}
\newpage
\baselineskip 16pt plus 2pt minus 2pt

\thispagestyle{empty}

\par
\topmargin=-1cm      

{ \small

\hfill{DOE/ER/40762-164}

\hfill{UMDPP\#99-039}

\hfill{NT@UW-99-7}

\hfill{KRL MAP-239}}

\vspace{20.0pt}

\begin{centering}
{\Large\bf Compton Scattering on the Deuteron in\\
Baryon Chiral Perturbation Theory}\\

\vspace{30.0pt}
{{\bf S.R.~Beane}$^1$,
{\bf M.~Malheiro}$^{2,1}$,\\
{\bf D.R.~Phillips}$^{3,1}$ and
{\bf U.~van Kolck}$^{4,3}$ }\\
\vspace{20.0pt}
{\sl $^{1}$Department of Physics,} \\ 
{\sl University of Maryland,
College Park, MD 20742}\\
{\tt sbeane@physics.umd.edu}\\  
\vspace{15.0pt}
{\sl $^{2}$Instituto de F\'{\i}sica, } \\
{\sl Universidade Federal Fluminense, 24210-340, Niter\'oi, R.J., Brazil } \\
{\tt mane@if.uff.br}\\
\vspace{15.0pt}
{\sl $^{3}$Department of Physics}\\
{\sl University of Washington, Seattle, WA 98195-1560} \\
{\tt phillips@dirac.phys.washington.edu}\\
\vspace{15.0pt}
{\sl $^{4}$Kellogg Radiation Laboratory, 106-38} \\
{\sl California Institute of Technology, Pasadena, CA 91125}\\
{\tt vankolck@krl.caltech.edu}\\
\end{centering}
\vspace{20.0pt}

\begin{center}
\begin{abstract}
Compton scattering on the deuteron is studied in the framework of
baryon chiral perturbation theory to third order in small momenta, for
photon energies of order the pion mass. The scattering amplitude is a
sum of one- and two-nucleon mechanisms with no undetermined
parameters.  Our results are in good agreement with existing
experimental data, and a prediction is made for higher-energy data
being analyzed at SAL.
\end{abstract}

\vspace*{10pt}

\vfill
\end{center}

\newpage

\section{Introduction}
\label{sec-intro}

Experimental facilities which accurately measure the energy of a
photon beam using photon tagging have made possible a new generation
of experiments which probe the low-energy structure of nucleons and
nuclei.  In particular, photon tagging can be used to measure Compton
scattering on weakly-bound systems, since it facilitates the
separation of elastic and inelastic cross sections. At sufficiently
low energy $\omega$ the spin-averaged forward Compton scattering
amplitude for any nucleus is, in the nuclear rest frame:

\begin{equation}
T(\omega)=-\frac{1}{4\pi} \vepsprime \cdot \veps
\left(\frac{{\cal Z}^2 e^2}{A M} + (\alpha + \beta) \omega^2 + \ldots \right),
\end{equation}
where $\veps$ and $\vepsprime$ are the polarization vectors of the
initial and final-state photons, ${\cal Z} e$ is the total charge of
the nucleus, and $A M$ is its total mass. The first term in this
series is a consequence of gauge invariance, and is the Thomson limit
for low-energy scattering on a target of mass number $A$ and atomic
number ${\cal Z}$. The coefficient of the second term is the sum of
the target electric and magnetic polarizabilities, $\alpha$ and
$\beta$, respectively. The polarizabilities can be separated by going
to backward angles, which probes the difference, $\alpha -\beta$. The
polarizabilities contain information about the electromagnetic
structure of the target.

A number of recent experiments have focused on Compton scattering on
the proton and on the deuteron.  These experiments have already
resulted in a wealth of new data on $\alpha_p$ and $\beta_p$, the
proton electric and magnetic polarizabilities, and might yield
information on $\alpha_n$ and $\beta_n$, the corresponding neutron
quantities.  These are basic low-energy hadron parameters which
reflect the underlying QCD dynamics. In a simple quark model picture
they contain averaged information about the charge and current
distribution produced by the quarks inside the nucleon.

At energies well below the chiral symmetry breaking scale, $\lsc\sim 4
\pi f_\pi \sim {M} \sim {m_\rho}$, the electromagnetic interactions of
pions and nucleons can be described systematically using an effective
field theory.  This effective field theory, known as chiral
perturbation theory ($\cpt \,$), reflects the observed QCD pattern of
symmetry breaking; there is an approximate $SU(2)_L \times SU(2)_R$
symmetry which is spontaneously broken to $SU(2)_V$ (isospin).  This
pattern of spontaneous symmetry breaking gives rise to three Goldstone
bosons which are identified with the pions. In the chiral limit pions
interact only through derivative interactions. A finite pion mass
reflects the nonvanishing current quark masses in QCD.  $\cpt$ is a
useful theory because pions interact only through vertices with a
countable number of derivatives and/or insertions of the quark mass
matrix; each power of the pion mass and each derivative count as one
power of ``small momenta''. Therefore $S$-matrix elements computed in
$\cpt$ can be expressed as simultaneous expansions in powers of
momenta and pion masses over the characteristic scale of physics that
is not included explicitly in the effective theory:

\begin{equation}
\frac{Q}{\Lambda_\chi},
\end{equation}
taking $Q$ to represent a ``small momentum'': a momentum $p \ll \lsc$
or a pion mass.  Note that the dynamical effects of {\it all} mesons
other than the pion are accounted for through local pionic operators.
In the purest form of $\cpt$, the coefficients of these operators are
fit to one experiment and then used to predict other experiments.
Hence $\cpt$ is a predictive theory.  The machinery of $\cpt$ was
originally developed for Goldstone boson interactions. However, it is
straightforward to extend the machinery to include single-baryon
processes, with the baryons treated as heavy, static objects. This
subject is reviewed in great detail in Ref.~\cite{bkmrev}.

Nucleon Compton scattering has been studied in $\cpt$ in
Ref.~\cite{ulf1}, where the following results for the polarizabilities
were obtained to order $Q^3$:

\begin{eqnarray}
\alpha_p=\alpha_n=\frac{5 e^2 g_A^2}{384 \pi^2 f_\pi^2 m_\pi} &=&12.2 \times
10^{-4} \, {\rm fm}^3; \label{eq:alphaOQ3}\nonumber\\ 
\beta_p=\beta_n=\frac{e^2 g_A^2}{768 \pi^2 f_\pi^2 m_\pi}&=& 1.2 \times 
10^{-4} \, {\rm fm}^3. \label{eq:betaOQ3}
\end{eqnarray}
Here we have used $g_A=1.26$ for the axial coupling of the nucleon,
and $f_\pi=93$ MeV as the pion decay constant. Note that the
polarizabilities are {\it predictions} of $\cpt$ at this order.  The
$O(Q^3)$ $\cpt$ predictions diverge in the chiral limit because they
arise from pion loop effects. In less precise language, the power
counting of $\cpt$ implies that polarizabilities are dominated by the
dynamics of the long-ranged pion cloud surrounding the nucleon, rather
than by short-range dynamics.  The polarizabilities should thus
provide a sensitive test of chiral dynamics.

At the next order in the chiral expansion, $Q^4$, there are
contributions to the polarizabilities from undetermined parameters
which must be fixed independently~\cite{ulf2}.  These counterterms
account for short-range contributions to the nucleon structure.  This
higher-order calculation will be discussed further below.

Recent experimental values for the proton polarizabilities are
\cite{newanalysis} \footnote{These are the result of a model-dependent
  fit to data from Compton scattering on the proton at several angles
  and at energies ranging from 33 to 309 MeV.}

\begin{eqnarray}
\alpha_p + \beta_p=13.23 \pm 0.86^{+0.20}_{-0.49} \times 10^{-4} \, {\rm fm}^3,
\nonumber\\
\alpha_p - \beta_p=10.11 \pm 1.74^{+1.22}_{-0.86} \times 10^{-4} \, {\rm fm}^3,
\label{protpolexpt}
\end{eqnarray}
where the first error is a combined statistical and systematic error,
and the second set of errors comes from the theoretical model
employed. We note that the value of $\alpha_p + \beta_p$ extracted
from this multi-energy fit is consistent with the venerable Baldin sum
rule~\cite{baldin}. These values are also in good agreement with the
chiral perturbation theory predictions.  The neutron polarizabilities
are difficult to obtain experimentally and so the corresponding $\cpt$
prediction is not well tested.  A dispersion sum rule can relate the
sum of electric and magnetic polarizabilities to an integral of the
deuteron photo-absorption cross section; with model-dependent
assumptions about the size of the neutron contribution to the deuteron
photo-absorption cross section, it has been found that~\cite{baldin}

\begin{equation}
\alpha_n + \beta_n=(14.40 \pm 0.66) \times 10^{-4} \, {\rm fm}^3.
\label{neutpolexpt1}
\end{equation}
Most direct information on $\alpha_n$ has been obtained by scattering
neutrons on a heavy nucleus and examining the cross section as a
function of energy. Currently there is much controversy over what
result this technique gives for $\alpha_n$.  In Ref. \cite{neutpol1}
the value~\footnote{The values measured in these experiments are {\it
    static} polarizabilities of the neutron, rather than the
   polarizabilities (often denoted $\bar{\alpha}$) we
  have been discussing here. In order to correct for this we have
  applied the standard 5\% correction to obtain $\bar{\alpha}$'s
  from the experimental results~\cite{jerry}.}

\begin{equation}
\alpha_n=(12.6 \pm 1.5 \pm 2.0)  \times 10^{-4} \, {\rm fm}^3
\label{neutpolexpt2}
\end{equation}
was obtained, which disagrees considerably with the result of the 
experiment of Ref. \cite{neutpol2},
\begin{equation}
\alpha_n=(0.6 \pm 5.0) \times 10^{-4} \, {\rm fm}^3 .
\label{neutpolexpt3}
\end{equation}

Given this discrepancy, it is important to measure the electromagnetic
neutron polarizabilities by other means.  Compton scattering on a
nuclear target is an obvious candidate.  Quasi-free Compton scattering
by the neutron in a deuteron target was reported in
Ref. \cite{neutpol3}.  With the use of a theoretical model, the
polarizability

\begin{equation}
\alpha_n=(10.7^{+ 3.3}_{- 10.7})  \times 10^{-4} \, {\rm fm}^3
\label{neutpolexpt4}
\end{equation} 
was extracted. One difficulty with this extraction of the
polarizability by scattering on a nuclear target lies in the absence
of a large neutron contribution: for quasi-free scattering the cross
section in the forward direction goes as $\sigma_n \sim (\alpha_n
\omega^2)^2$ (where we have neglected the smaller magnetic
polarizability) and is very small. Partly for this reason, attention
has turned to other methods of extracting the neutron
polarizabilities.  It should be pointed out though that some authors
argue that quasi-free photon-neutron scattering deserves renewed
study~\cite{wissmann}.

One alternative to this quasi-free Compton scattering is coherent
photon scattering on the deuteron. In this case, the cross section in
the forward direction naively goes as:

\begin{equation}
\left.\frac{d \sigma}{d \Omega} \right|_{\theta=0}
\sim (f_{Th} - (\alpha_p + \alpha_n) \omega^2)^2.
\end{equation} 
The sum $\alpha_p + \alpha_n$ may then be accessible via its
interference with the dominant Thomson term for the proton,
$f_{Th}$~\cite{hornidge}. This is one example of a general argument
that the deuteron differential cross section is sensitive to isoscalar
combinations of the nucleon polarizabilities.  This means that with
experimental knowledge of the proton polarizabilities it may be
possible to extract those for the neutron.  Coherent Compton
scattering on a deuteron target has been measured at $E_\gamma=$ 49
and 69 MeV by the Illinois group \cite{lucas}.  An experiment with
tagged photons in the energy range $E_\gamma= 84.2-104.5$ MeV is under
analysis at Saskatoon \cite{SAL}, while data for $E_\gamma$ of about
60 MeV is being analyzed at Lund~\cite{lund}.

There is clearly a need for a consistent theoretical framework in
order to interpret this data. The amplitude for Compton scattering on
the deuteron involves mechanisms other than Compton scattering on the
individual constituent nucleons.  Hence, extraction of nucleon
polarizabilities requires a theoretical calculation of Compton
scattering on the deuteron that is under control in the sense that it
accounts for {\it all} mechanisms to a given order in a systematic
expansion in a small parameter.  There exist a few calculations of
this reaction in the framework of conventional potential models
\cite{wilbois,levchuk,jerry}.  These calculations yield similar
results if similar input is supplied, but typically mechanisms for
nucleon polarizabilities and two-nucleon contributions are not treated
consistently.  We will see that $\cpt$ provides an alternative
framework where this drawback can be eliminated.

In few-nucleon systems, a complication arises in $\cpt$ due to the
existence of shallow nuclear bound states and related infrared
singularities in $A$-nucleon reducible Feynman diagrams evaluated in
the static approximation~\cite{weinnp}. Physically, the fundamental
problem is that nuclear physics introduces a new mass scale, the
nuclear binding energy, which is very small compared to a typical
hadronic scale.  One way to overcome this difficulty is to adopt a
modified power counting scheme in which $\cpt$ is used to calculate an
effective potential which generally consists of all $A$-nucleon
irreducible graphs. The $S$-matrix, which includes all reducible
graphs as well, is then obtained through iteration by solving a
Lippmann-Schwinger equation~\cite{weinnp}. (We will refer to
this version of the effective theory as the Weinberg formulation.)  To
date the Weinberg formulation can be carried through rigorously only
using finite cutoff regularization~\cite{ordonez,vk,ray,lepage}.  This
limitation has spawned an intense theoretical effort geared at
formulating an effective field theory for low-lying bound states which
is verifiably (analytically) consistent in the sense of
renormalization~\cite{seki}. One result of this effort is a new power
counting scheme in which all nonperturbative physics responsible for
the presence of low-lying bound states arises from the iteration of a
single operator in the effective theory, while all other effects,
including all higher dimensional operators {\it and} pion exchange,
are treated perturbatively~\cite{pc,ksw}. (We will refer to this
version of the effective theory as the Kaplan-Savage-Wise (KSW)
formulation.) This is relevant here because Compton scattering on the
deuteron has been computed to next-to-leading order in the KSW
formulation~\cite{martinetal}.  We will discuss this result and its
relation to our calculation.  A comprehensive and up-to-date review of
nuclear applications of effective field theories can be found in
Ref.~\cite{monster}.

\begin{figure}[t,h,b,p]
   \vspace{0.5cm}
   \epsfysize=6.2cm
   \centerline{\epsffile{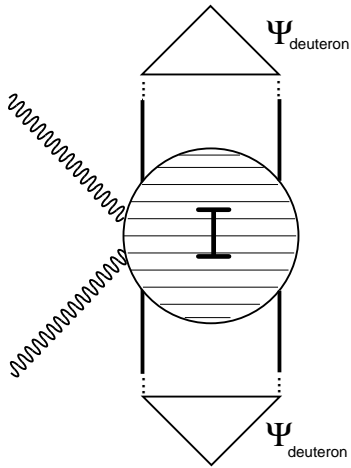}}
   \centerline{\parbox{11cm}{\caption{\label{fig5}
The anatomy of the calculation. The irreducible kernel (I) is computed
in baryon $\cpt$ and sewn to external deuteron wavefunctions to give
the matrix element for Compton scattering on the deuteron.
  }}}
\end{figure}

In this paper we compute Compton scattering on the deuteron for
incoming photon energies of order 100 MeV in the Weinberg
formulation. We use baryon $\cpt$ to compute an irreducible scattering
kernel to order $Q^3$, which is then sewn to external deuteron
wavefunctions as in Fig. \ref{fig5}.  Compton scattering on the
nucleon is a fundamental ingredient of our calculation. This is a
direct consequence of the power counting scheme specific to the
Weinberg formulation.  It is important to realize that typical nucleon
momenta inside the deuteron are small---on the order of $\sqrt{MB}$ or
$m_\pi$, with $B$ the deuteron binding energy---and consequently, {\it
a priori} we expect no convergence problems in the $\cpt$ expansion of
any low-momentum electromagnetic or pionic probe of the deuteron.
This method has led to fruitful computation of the pion-deuteron
scattering length~\cite{weinnp,silas1}, neutral pion photoproduction
on the deuteron at threshold~\cite{silas2}, as well as $pn$ radiative
capture~\cite{rho1} and the solar burning process $p p \rightarrow d
e^+ \nu$~\cite{rho2}. Although in principle we could use wavefunctions
computed in $\cpt$, we will consider wavefunctions generated using two
modern nucleon-nucleon potentials. In previous computations using this
method results are not very sensitive to details of the deuteron
wavefunction.  Any wavefunction with the correct binding energy gives
equivalent results to within the theoretical error expected from
neglected higher orders in the chiral expansion. Similar wavefunction
independence is found here.

In Section~\ref{sec-bcpt} we review those aspects of $\cpt$ that are
directly relevant to our calculation, including the power counting
scheme and the relevant operators.  In Section~\ref{sec-counting} we
present the power counting for Compton scattering on the $NN$
system. We discuss subtleties which arise because of the presence of a
new mass scale in the two-nucleon system: the deuteron binding
energy. Then, in Section~\ref{sec-calc} we present our calculation of
Compton scattering on the $NN$ system to order $Q^3$ in baryon
$\cpt$. We discuss Compton scattering on a single nucleon---which has
been calculated elsewhere \cite{bkmrev,ulf1,ulf2}---and compute the
Feynman amplitudes which contribute to photon scattering on the
two-nucleon system at order $Q^3$ in the momentum expansion. In
Section~\ref{sec-results} we present results for differential
cross sections for Compton scattering on unpolarized deuteron targets.
We discuss higher-order effects and the extent to which a
determination of neutron polarizabilities is possible.  Finally, in
Section~\ref{sec-conclusion} we summarize our work, and identify
future directions for the theoretical study of this reaction.

\section{Baryon Chiral Perturbation Theory}
\label{sec-bcpt}

In this section, we briefly discuss the effective chiral Lagrangian
underlying our calculations and the corresponding power counting.
Precise statements about QCD dynamics at low energies can be made only
where a small dimensionless expansion parameter is identified. This is
of course the motivation behind the ongoing intense effort to develop
a perturbative theory of nuclear interactions~\cite{monster}.  One
class of dimensionless parameters consists of pure numbers, such as a
coupling constant associated with a renormalizable interaction, or the
inverse of the dimensionality of a group, as in the large-$N_{c}$
expansion in QCD. A second class of dimensionless parameters consists
of ratios of dimensional quantities which are present in a physical
system when there is a clean hierarchy of scales. Effective field
theory is the technology which develops a hierarchy of scales into a
perturbative expansion of physical observables. In a system with
broken symmetries this technology is especially powerful. When a
continuous symmetry is spontaneously broken there are always massless
Goldstone modes which dominate the low-energy dynamics. Although the
symmetry in question is no longer a symmetry of the vacuum, it remains
a symmetry of the Lagrangian. Moreover, massless Goldstone modes only
couple derivatively. Hence at energies small relative to the
characteristic symmetry breaking scale, the interactions of the
Goldstone bosons can be ordered in an effective Lagrangian which is
constrained by chiral symmetry and in which each operator contains a
nonvanishing number of derivatives acting on the pion fields.
Observables computed from the effective Lagrangian are therefore power
series in momenta, with the non-analyticities required by perturbative
unitarity.

In QCD the chiral $SU(2)_L \times SU(2)_R$ symmetry is spontaneously
broken and the low-energy effective field theory is $\cpt$. Here we
are interested in processes where the typical momenta of all external
particles is $p\ll\lsc$, so we identify our expansion parameter as
$p/\lsc$. If the chiral symmetry of QCD were exact this would be the
only expansion parameter in $\cpt$.  However, in QCD $SU(2)_L \times
SU(2)_R$ is softly broken by the small quark masses. This explicit
breaking implies that the pion has a small mass in the low-energy
theory.  Since ${m_\pi}/\lsc$ is then also a small parameter, we have a
dual expansion in $p/\lsc$ and ${m_\pi}/\lsc$. We take $Q$ to
represent either a small momentum {\it or} a pion mass. A generic
contribution to a matrix element involving the interaction of any
number of pions and $A$ nucleons can then be written in the form

\begin{equation}
{\cal M}={Q^\nu}{\cal F}(Q/\mu),
\end{equation}
where $\mu$ is a renormalization scale, ${\cal F}(x)$ is some
non-analytic function, which is assumed to obey ${\cal F}(1)=O(1)$,
and $\nu$ is a counting index.  In general the full ${\cal M}$ is the
sum of a set of Feynman diagrams in a quantum field theory.  It is
straightforward to arrive at a general formula for $\nu$ by
considering the momentum-space structure of generic Feynman
rules~\cite{weinnp}.

First, we consider only irreducible subdiagrams. These are graphs in
which the energies flowing through all internal lines are of $O(Q)$.
The issue of calculating the full matrix element for the process of
interest from this kernel will be taken up shortly.
In irreducible diagrams we have the following counting rules:

\begin{itemize}
\item A nucleon propagator contributes $Q^{-1}$;

\item A pion propagator contributes $Q^{-2}$;

\item Each derivative or power of the pion mass at a vertex
contributes $Q$;

\item Each loop integral contributes $Q^4$;

\item Each delta function of four-momentum conservation contributes $Q^{-4}$.
\end{itemize}

\noindent
In this way, we find that a diagram with $A$ nucleons, $L$ loops, $C$
separately connected pieces, and $V_i$ vertices of type $i$
gives~\cite{weinnp,Friar}

\begin{equation}
\nu = 4-{A}-2C+2L+{\sum _i}{V_i}{\Delta _i},
\label{nu}
\end{equation}
where the so-called index of the interaction $i$ is

\begin{equation}
{\Delta _i}\equiv {d_i}+{f_i}/2-2
\label{Deltai}
\end{equation}
with $d_i$ the sum of the number of derivatives or powers of $m_\pi$ and 
$f_i$ the number of nucleon fields.

Since pions couple to each other and to nucleons through either
derivative interactions or quark masses, there is a lower bound on the
index of the interaction: ${\Delta _i}\geq 0$. Hence the leading {\it
irreducible} graphs are tree graphs ($L=0$) with the maximum number of
separately connected pieces ($C=max$), and with vertices with $\Delta
_i =0$.  Higher-order graphs (with larger powers of $\nu$) are
perturbative corrections to this leading effect. In this way, a
perturbative series is generated by increasing the number of loops
$L$, decreasing the number of connected pieces $C$, and increasing the
number of derivatives, pion masses and/or nucleon fields in
interactions.

How is this analysis altered in the presence of an electromagnetic
field? Photons couple via the electromagnetic field strength tensor
and by minimal substitution. This has the simple effect of replacing a
derivative by a factor of the charge $e$.  This introduces a second
expansion, in the small electromagnetic coupling $\alpha_{\rm
em}=e^2/4\pi$.  Since we will be working at a fixed order in the
expansion in $\alpha_{\rm em}$ (leading order), it is convenient to
enlarge the definition of $d_i$ to include powers of $e$ as well, thus
continuing to classify interactions according to the index $\Delta _i$
defined by (\ref{Deltai}).

With the power counting scheme established, the next step is to
construct the effective Lagrangians labelled by the interaction index
$\Delta_i$.  The technology that goes into building an effective
Lagrangian and extracting the Feynman rules is standard by now and
presented in great detail in \cite{bkmrev}.

The pion triplet is contained in a matrix field

\begin{equation}
\Sigma =\xi^2 \equiv \sqrt{1-\frac{\vec\pi^2}{f_\pi^2}} 
        +i \frac{{\vec\pi}\cdot{\vec\tau}}{f_\pi},
\end{equation}
where $f_\pi =93$ MeV is the pion decay constant.  Under
$SU(2)_L\times SU(2)_R$, $\Sigma$ transforms as $\Sigma\rightarrow
L\Sigma{R^\dagger}$ and $\xi$ as $\xi\rightarrow
L\xi{U^\dagger}=U\xi{R^\dagger}$; here $L$($R$) is an element of
$SU(2)_L$ ($SU(2)_R$), and $U$ is defined implicitly.  It is
convenient to assign the nucleon doublet $N$ the transformation
property $N\rightarrow UN$.

With $Q=(1+{\tau _3})/2$ we can write the pion covariant derivative as

\begin{equation}
{D_\mu}\Sigma={\partial _\mu}\Sigma -ie{{\cal A}_\mu}[Q,\Sigma].
\end{equation}
Out of $\xi$ one can construct

\begin{eqnarray}
{V_\mu}&=&\frac{1}{2}[{\xi ^\dagger}({\partial _\mu}-ie{{\cal A}_\mu}Q)\xi +
 \xi({\partial _\mu}-ie{{\cal A}_\mu}Q){\xi ^\dagger}] \\
{A_\mu}&=&\frac{i}{2}[{\xi ^\dagger}({\partial _\mu}-ie{{\cal A}_\mu}Q)\xi -
 \xi({\partial _\mu}-ie{{\cal A}_\mu}Q){\xi ^\dagger}], 
\end{eqnarray}
which transform as ${V_\mu}\rightarrow U{V_\mu}{U^\dagger} +
U{\partial _\mu}{U^\dagger}$ and ${A_\mu}\rightarrow
U{A_\mu}{U^\dagger}$ under $SU(2)_L\times SU(2)_R$.  $V_\mu$ is used
to build covariant derivatives of the nucleon,

\begin{equation}
D_\mu N= ({\partial _\mu}+{V_\mu})N.
\end{equation}
The electromagnetic field enters not only through minimal coupling in
the covariant derivatives, but also through

\begin{equation}
f_{\mu\nu}= e({\xi ^\dagger}Q\xi + \xi Q{\xi ^\dagger}){F_{\mu\nu}},
\end{equation}  
where ${F_{\mu\nu}}$ is the usual electromagnetic field strength
tensor.

Because the nucleon mass is large, $M \gg Q$, it plays no dynamical
role: nucleons are non-relativistic objects in the processes we are
interested in. The field $N$ can be treated as a heavy field of
velocity $v$ in which on-mass-shell propagation through $\exp
(iM{v\cdot x})$ has been factored out.  In the rest frame of the
nucleon, the velocity vector ${v^\mu}=(1,\vec{0})$.  The spin operator
is denoted by $S^\mu$, and in the nucleon rest frame
$S^\mu=(1/2)(0,\vec{\sigma})$.

With these ingredients, it is straightforward to construct the
effective Lagrangian, conveniently ordered according to the index
$\Delta_i$ of Eq. (\ref{Deltai}),

\begin{equation}
{\cal L} =\sum_{\Delta=0}^{\infty} {\cal L}^{(\Delta)}.
\end{equation}
The leading-order Lagrangian is

\begin{eqnarray}
{\cal L}^{(0)}& = &
     \frac{1}{4}{f_\pi^2}tr({D_\mu}{\Sigma ^\dagger}{D^\mu}{\Sigma})
    +\frac{1}{4}{{f_\pi^2}{m_\pi^2}}tr({\Sigma}+{\Sigma ^\dagger}) \nonumber \\
& & +i N^\dagger (v\cdot D) N
               + 2g_A N^\dagger (A\cdot S)N \nonumber \\
 & & -\frac{1}{2}{C_a}(N^\dagger {\Gamma _a}N)^2
\label{L0}
\end{eqnarray}
where $\Gamma _a$ is an arbitrary (non-derivative) Hermitian operator,
and $g_A=1.26$ and $C_a$ are parameters.  The sub-leading terms
are contained in

\begin{eqnarray}
{\cal L}^{(1)}& = &
\frac{1}{2M}N^\dagger({-D^2}+(v\cdot D)^2
                    +2i{g_A}\{{v\cdot A},{S\cdot D}\} \nonumber \\           
& &-\frac{i}{2}[{S^\mu},{S^\nu}][(1+{\kappa _v}){f_{\mu\nu}}
+ \frac{1}{2}({\kappa_s}-{\kappa_v})tr{f_{\mu\nu}}])N +\ldots,
\label{L1}
\end{eqnarray}
where ${\kappa _v}={\kappa _p}-{\kappa _n}$ and ${\kappa _s}={\kappa
_p}+{\kappa _n}$ are parameters related to the anomalous magnetic
moments of the proton, $\kappa _p=1.79$, and neutron, $\kappa
_n=-1.91$. In sub-subleading order,

\begin{equation}
{\cal L}^{(2)} = -\frac{e^2}{32 \pi^2 f_\pi} \pi_3
                  \epsilon^{\mu\nu\rho\sigma}F_{\mu\nu}F_{\rho\sigma}
+ {\frac{e^2}{4{M^2}}}{{\cal Z}({\cal Z}+2\kappa )}N^\dagger 
\vec\sigma\cdot ({\vec A}\times{\vec E})N +\ldots
\label{L2}
\end{equation}
where the first term is from the anomaly, with $\epsilon_{0123}=1$,
and the second term is a $1/{M^2}$ correction expressed in the nucleon
rest frame with electric field ${\vec E}={\partial_0}{\vec
A}$~\cite{ulf1}. In our notation, $\kappa={\kappa _p}$ when ${\cal
Z}=1$ and $\kappa={\kappa _n}$ when ${\cal Z}=0$ and in the
expressions above, the ellipses signify that we have included only the
interactions which are relevant to our calculation.  Likewise, terms
in ${\cal L}^{(n\ge 3)}$ do not appear in Compton scattering to the
order that we are working.

Note, finally, that the $\Delta$ isobar can be introduced in a manner
analogous to the nucleon field.  The delta-nucleon mass difference
$M_\Delta-M$ is not much larger than our $Q$, so integrating out the
delta may worsen the convergence of the low-energy expansion
considerably. Of course, the effects of the delta still appear in the
effective theory constructed above, via their inclusion in
counterterms, but keeping an explicit delta shifts these effects to
lower orders in the expansion. For simplicity, in this initial
calculation we do not include the delta.

\section{Compton Counting}
\label{sec-counting}

\subsection{The role of nuclear physics}

\begin{figure}[t,h,b,p]
   \vspace{0.5cm}
   \epsfysize=3.5cm
   \centerline{\epsffile{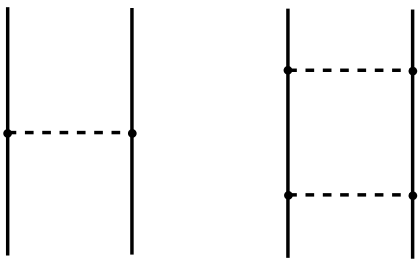}}
   \centerline{\parbox{11cm}{\caption{\label{fig3}
One- and two-pion exchange graphs. One-pion exchange is in the
potential and (uncorrelated) two-pion exchange is generated by
solving the Schr\"odinger equation.
  }}}
\end{figure}

The power counting of chiral perturbation theory summarized above
works very well for processes involving one or zero nucleons. A
non-trivial new element enters the effective theory when we consider
systems of more than one nucleon \cite{weinnp,monster}.  Because
nucleons are heavy, contributions from intermediate states that differ
from the initial state only in the kinetic energy of nucleons are
enhanced by infrared quasi-divergences. This is linked to the
existence of small energy denominators $\sim Q^2/M$, which generate
contributions $O(\Lambda_\chi/Q)$ larger than would be expected from
Eq.~(\ref{nu}). However, this is not really either a surprise or a
problem.  Eq.~(\ref{nu}) was derived by assuming that all nucleon
propagators scale as $1/Q$. This is correct for irreducible
subdiagrams, which do not contain intermediate states with small
energy denominators. In an $A$-nucleon system which is not subject to
any external probes these diagrams are $A$-nucleon irreducible
diagrams, the sum of which we call the potential $V$.  (See e.g. the
left-hand graph of Fig.~\ref{fig3}.)  To obtain the full two-nucleon
Green's function these irreducible graphs are iterated using the free
two-nucleon Green's function, which can, of course, have a small
energy denominator which does not obey the $\cpt$ power-counting. This
iteration gives contributions to the two-nucleon Green's function such
as the graph shown on the right-hand side of Fig.~\ref{fig3}.  When we
consider external probes with momenta of order $Q$, the sum of
irreducible diagrams forms a kernel $K$ for the process of
interest. The full Green's function for the reaction is then found by
multiplying this kernel $K$ by two-particle Green's functions in which
the two-particle propagators with small energy denominators may
appear.

In the case of Compton scattering on the two-nucleon system the
formal version of these statements is as follows. The
two-particle Green's function, $G$, is constructed from the
two-nucleon-irreducible interaction $V$, and the free
two-nucleon Green's function $G_0$, via

\begin{equation}
G=G_0 + G_0 V G.
\end{equation}
Since $V$ is two-particle irreducible we may perform a chiral power
counting on it as per Eq. (\ref{nu}). Such a potential has been
constructed up to $\nu =3$, in Refs.~\cite{ordonez,vk}. That work
employed cutoff regularization, and the Schr\"odinger equation was
solved to obtain the $NN$ wave function.  Fits to deuteron properties
and scattering data up to laboratory energies of 100 MeV were obtained
for three different cutoff parameters~\cite{ray}. However, the quality
of the fit is not as good as that of phenomenological models. Full
consistency can be achieved by treating both the kernel for the
interaction with external probes and the $NN$ potential in the same
framework. However, here we will follow a hybrid approach: we will
calculate the kernel in $\cpt$ but use several ``realistic'' potential
models to calculate the two-particle Green's function.

To calculate Compton scattering on the two-nucleon system we must also
identify the two-nucleon irreducible kernel, $K_{\gamma \gamma}$, for
the process $\gamma NN \rightarrow \gamma NN$.  Again, the
power-counting of Eq.~(\ref{nu}) applies to this object, since all
nucleon denominators in it are, by definition, of $O(Q)$, as long as
$\omega\sim Q$. The full Green's function for Compton scattering on
the two-nucleon system is then

\begin{equation}
G_{\gamma \gamma}=G K_{\gamma \gamma} G.
\label{eq:ComptonGF}
\end{equation}
It follows that if $K_{\gamma \gamma}$ is constructed in a gauge
invariant way and is worked out to the same order in $\cpt$ as the
potential $V$ then this calculation will be gauge invariant up to that
order in chiral perturbation theory.  However, even if
$K_{\gamma\gamma}$ and $V$ {\it are} consistently constructed this
calculation is not {\it exactly} gauge invariant, since a brief
examination of the counting explained above shows that contributions
to $K_{\gamma \gamma}$ of the form

\begin{equation}
K_\gamma G K_\gamma,
\label{eq:resonance}
\end{equation}
where $K_\gamma$ is the two-nucleon-irreducible kernel for one photon
interacting with the two-nucleon system, only appear order-by-order in
the expansion in $Q/\lsc$. In fact, for gauge invariance to be {\it
exact} the contribution (\ref{eq:resonance}) must be included in full
in $G_{\gamma \gamma}$.

Indeed, at low photon energies the contribution (\ref{eq:resonance}),
known as the resonance contribution to Compton scattering, is crucial
to a correct understanding of the reaction. In particular, without it
the Thomson limit amplitude for Compton scattering on the deuteron
will not be recovered. In this low-energy region the power-counting of
Eq.~(\ref{nu}) does not apply to $K_{\gamma \gamma}$, since graphs in
$K_{\gamma \gamma}$ with two-nucleon intermediate states
(Eq. (\ref{eq:resonance})) can lead to violations of the power-counting
via the same infrared quasi-divergences discussed above. However, if
the energy of the probe is larger ($\sim {m_\pi}$) it is valid to look
at the contributions $K_\gamma G K_\gamma$ using the power-counting of
Eq.~(\ref{nu}). Hence at higher photon energies it is completely
correct to treat this resonance contribution only perturbatively in
the $Q$ expansion.  In other words, as we move into the energy region
which is well above the poles of the Green's function $G$, it becomes
reasonable to use a perturbative expansion for $G$ in
Eq.~(\ref{eq:resonance}), rather than one which retains its full
structure.

Arguments such as these lead us to identify two regimes for Compton
scattering on nuclei, depending on the relation of the energy of the
photonic probe, $\omega$, to the typical nuclear binding scale $B\sim
m_\pi^2/M$:

\begin{enumerate}
\item $\omega \sim B$. In this region the power-counting of Eq.~(\ref{nu})
is not valid, since the external probe momentum flowing through the
nucleon lines is of order $Q^2/M$, rather than order $Q$. It is in
this region that the Compton low-energy theorems are derived. Therefore
our power counting will not recover those low-energy theorems.

\item $\omega \sim Q \gg B$. In this region the power counting of
Eq.~(\ref{nu}) is valid, since in all irreducible graphs a
photon energy of order $Q$ flows through the nucleon lines. 
\end{enumerate}

\begin{figure}[t,h,b,p]
   \vspace{0.5cm} \epsfysize=5.5cm
   \centerline{\epsffile{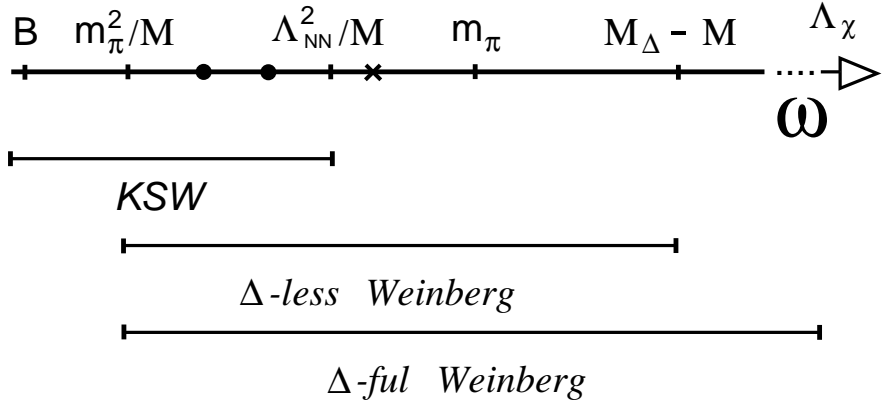}}
   \centerline{\parbox{11cm}{\caption{\label{efftheory} Hierarchy of
   scales and effective field theories of Compton scattering as a
   function of photon energy $\omega$, as discussed in the text.  The
   dots represent the experimental data from Illinois and the cross
   represents the forthcoming data from Saskatoon.}}}
\end{figure}

These two regimes were identified by Friar and Tomusiak in a
discussion of the role of low-energy theorems for threshold pion
photoproduction on nuclei~\cite{FT81}. They were also discussed in the
work of Chen {\it et al.}~\cite{martinetal}. In Ref.~\cite{martinetal}
Compton scattering on the deuteron was computed to the same order
discussed here, one order beyond leading non-vanishing order. However,
there the calculation was performed in the KSW formulation of
two-nucleon effective field theory, rather than in the Weinberg
formulation.  An advantage of KSW power-counting is that the effective
field theory moves smoothly between $Q < B$ and $Q > B$.  KSW
power-counting is valid for nucleon momenta $Q< {\Lambda_{NN}}\sim
300$ MeV. Thus in the KSW formulation deuteron polarizabilities and
Compton scattering up to energies $\omega < {\Lambda_{NN}^2}/{M}\sim
90$ MeV can be discussed in the same framework.  Here we are
interested mostly in the region $\omega\sim m_\pi$, and so we regard
ourselves as being firmly in the second regime ---although we shall
see that this requires us to be cautious in our conclusions for
Compton scattering at lower energies. In Fig.~\ref{efftheory} we
illustrate the ranges of validity of the KSW and Weinberg formulations
of effective field theory of Compton scattering on the deuteron.

\subsection{Power counting for $\gamma NN \rightarrow \gamma NN$}

\label{sec-kernelcounting}

Let us now consider in more detail the kernel in the regime $\omega
\sim m_\pi$.  By definition, it consists of those irreducible
subdiagrams to which the external photons are attached.  One cannot
cut these diagrams in two pieces without cutting at least one line
with energy $O(Q)$. In particular, there are no irreducible graphs
where the two photons are attached to different nucleons unless the
two nucleons are connected by some interaction: energy and momentum
conservation require the exchange of a pion or a short-range
interaction to carry the energy $O(Q)$ from one nucleon to the other,
and somewhere along the nucleon lines there is an energy flow of the
order of the pion mass.

We will concentrate here on real Compton scattering on an $A=2$ system
and work in the Coulomb gauge where the incoming and outgoing photon
momenta $k=(\omega, \vec{k})$ and $k'=(\omega', \vec{k}')$ and
polarization vectors $\epsilon$ and $\epsilon'$ satisfy:

\begin{equation}
k^2=0; \qquad k'^2=0; \qquad v \cdot \epsilon=0; \qquad v \cdot \epsilon' =0.
\end{equation}

\begin{center}
\underline{${\bf \nu =-2}$}
\end{center}
\noindent 

If we were interested in virtual photons, then the first contributions
would start at $\nu =-2$. The leading effect would be from tree graphs
($L=0$) with the maximum number of separately connected pieces ($C=2$)
constructed solely out of interactions with $\Delta _i =0$ (${\sum
_i}{V_i}{\Delta _i}=0$).  The only possible graphs of this type come
from the photon-nucleon interaction generated by minimal coupling in
${\cal L}^{(0)}$. For real photons these graphs vanish in the Coulomb
gauge.

\begin{center}
\underline{${\bf \nu =-1}$}
\end{center}
\noindent 

Likewise, some contributions at $\nu =-1$ are still tree graphs
($L=0$) with the maximum number of separately connected pieces
($C=2$), with one photon vertex of index one and one photon vertex of
index zero (${\sum _i}{V_i}{\Delta_i}=1$).  Graphs of this type do not
contribute to real Compton scattering, if the Coulomb gauge is used.
Similarly, all higher orders will include several diagrams that vanish
for real photons in Coulomb gauge. From now on we do not discuss such
graphs.

\begin{figure}[t]
   \vspace{0.5cm} \epsfysize=7.5cm
   \centerline{\epsffile{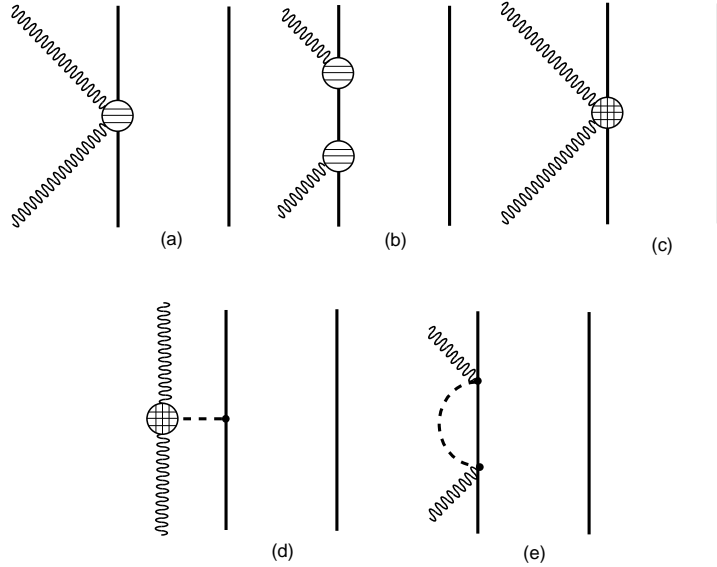}}
   \centerline{\parbox{11cm}{\caption{\label{fig1} Characteristic
   one-body interactions which contribute to Compton scattering on the
   deuteron at order $Q^2$ (a) and at order $Q^3$ (b-e) (in the
   Coulomb gauge). Crossed graphs are not shown and not all loop
   topologies are shown.  The sliced blobs are vertex insertions from
   ${\cal L}^{(1)}$. The sliced and diced blobs are vertex insertions
   from ${\cal L}^{(2)}$.  }}}
\end{figure}

Thus, at this order the only contribution comes from a tree-level
diagram with two separately connected pieces and a vertex which is
the two-photon seagull of $\Delta _i =1$. This vertex arises from
minimal coupling in the kinetic nucleon terms 
in ${\cal L}^{(1)}$ of Eq.~(\ref{L1}).  It is depicted in Fig.~\ref{fig1}(a).

\begin{center}
\underline{${\bf \nu =0}$}
\end{center}
\noindent 

\begin{figure}[t,h,b,p]
   \vspace{0.5cm}
   \epsfysize=7.5cm
   \centerline{\epsffile{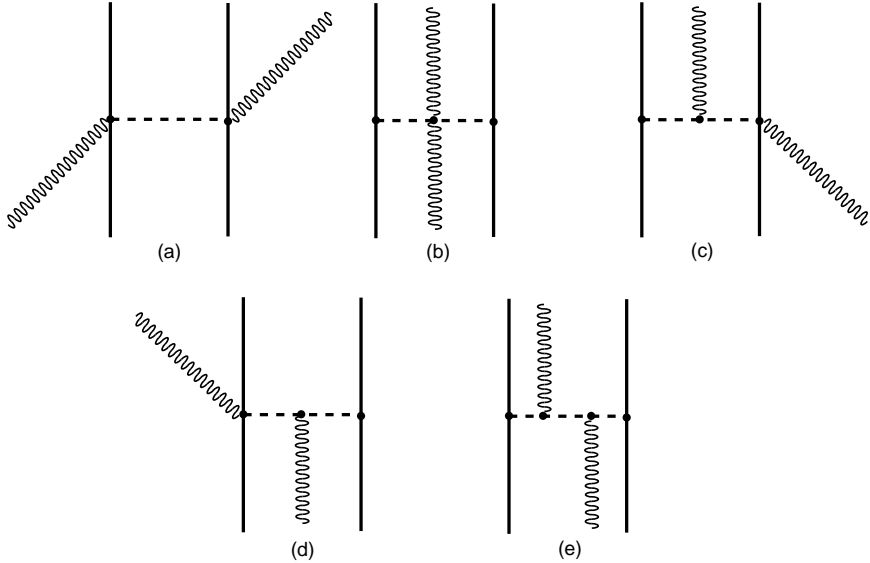}}
   \centerline{\parbox{11cm}{\caption{\label{fig2}
Two-body interactions which contribute to Compton scattering
on the deuteron at order $Q^3$. Permutations are not shown.
  }}}
\end{figure}

There are four types of corrections at this order:

\begin{enumerate}

\item Tree-level graphs ($L=0$) with two separately connected pieces
($C=2$) and two vertices with $\Delta _i =1$ (${\sum _i}{V_i}{\Delta
_i}=2$).  Interactions with $\Delta _i =1$ arise in ${\cal L}^{(1)}$
from minimal coupling in the kinetic nucleon term and from the Pauli
terms associated with the anomalous magnetic moments.  See
Fig. \ref{fig1}(b).

Note that if these calculations are performed in the $\gamma N N$
center-of-mass frame then the boost of the $\gamma N$ interaction from
the $\gamma N$ to the $\gamma NN$ c.m. frame appears naturally in the
calculation. It occurs as a modification to these $s$ and $u$-channel
pole graphs which must be included because the total $\gamma N$ system
three-momentum is non-zero.

\item Tree-level graphs ($L=0$) with two separately
connected pieces ($C=2$) and one interaction with $\Delta _i =2$
(${\sum _i}{V_i}{\Delta _i}=2$). Here a vertex 
in the ${\cal L}^{(2)}$ of Eq.~(\ref{L2})
which is a relativistic correction to the magnetic interaction
of the nucleon with the photons contributes. (See Fig. \ref{fig1}(c).)
There is also a process involving exchange of a $\pi^0$ with the
$\pi^0 \gamma \gamma$ interaction coming from the anomaly term in
${\cal L}^{(2)}$. (See Fig. \ref{fig1}(d).)

\item One-loop graphs ($L=1$) with two separately connected
pieces ($C=2$) and all interactions from the ${\cal L}^{(0)}$ 
of Eq.~(\ref{L0}) (${\sum _i}{V_i}\Delta _i =0$). (See Fig. \ref{fig1}(e).)

\item Tree graphs ($L=0$) with only one separately connected piece
($C=1$) and all interactions from ${\cal L}^{(0)}$
(See Fig. \ref{fig2}.)
\end{enumerate}

Note that we have not explicitly mentioned the delta anywhere
in this discussion of our power counting. In the energy range
of interest here, the delta could give important contributions. In
fact, with the delta ``integrated out'', as has been done here, its
contributions first appear at $\nu=1$, through counterterms.  If the
delta is included explicitly~\cite{jenkinsetal,hemmert}, then it enters at
order $\nu=0$ via the loop graphs discussed under 3 above.

\subsection{The breakdown of the power counting}

\begin{figure}[t,h,b,p]
   \vspace{0.5cm}
   \epsfysize=3.5cm
   \centerline{\epsffile{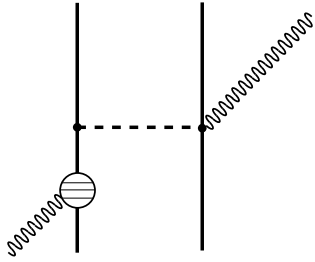}}
   \centerline{\parbox{11cm}{\caption{\label{fig4}
Two-body interaction which contributes to Compton
scattering on the deuteron at order $Q^4$. The sliced blob
represents a $1/M$ correction vertex from ${\cal L}^{(1)}$.
  }}}
\end{figure}

The identification of the relevant contributions can be extended to
higher orders in an obvious way.  An example of a $\nu = 1$ ($Q^4$)
contribution is shown in Fig. \ref{fig4}.  We can use this graph to
illustrate the transition to the very-low energy regime $Q\sim
m_\pi^2/M$.  It is easy to see that this graph becomes comparable to
the order $Q^3$ graph of Fig. \ref{fig2}(a) when

\begin{equation}
{\frac{|\vpee\; |^2}{\omega M}}\sim 1.
\end{equation}
Here $\vpee$ is a typical nucleon momentum inside the deuteron and
$\omega$ is the photon energy.  Since our power counting is predicated
on the assumption that all momenta are of order $m_\pi$, we find that
our power-counting formula (\ref{nu}) is valid for the kernel
$K_{\gamma\gamma}$ in the region

\begin{equation}
\frac{m_\pi^2}{M}\ll Q \ll \lsc .
\end{equation}

In this first paper we will limit ourselves to a $\nu =0$ ($Q^3$)
calculation in the deltaless theory. We plan to study effects of an
explicit delta field in a future publication.  It is remarkable that
to this order no unknown counterterms appear.  All contributions to
the kernel are fixed in terms of known pion and nucleon parameters
such as $m_\pi$, $g_A$, $M$, and $f_\pi$.  Thus, to this order $\cpt$
makes {\it predictions} for Compton scattering.

\section{The $\cpt$ calculation}
\label{sec-calc}

In this section, we outline how the various contributions are
calculated. We can separate the contributions discussed in the
previous section into two classes: one-nucleon and two-nucleon
mechanisms. One-nucleon diagrams are those which relate directly to
contributions to Compton scattering on a single nucleon, while
two-nucleon mechanisms contribute only in $A\ge 2$ nuclei. One-nucleon
scattering diagrams are traditionally labelled as $Q^n$ with $n$ the
index in the one-nucleon process.  We follow this usage here and above
to make closer contact with the one-nucleon calculations. Of course
only relative orders are physically significant.

\subsection{One-nucleon contribution up to order $Q^3$ ($\nu \le 0$)}

Let us consider first the one-nucleon terms; a sample of the relevant
diagrams is shown in Fig. \ref{fig1}.  In the $\gamma N$ c.m. frame
the nucleon amplitude can be written as

\begin{eqnarray}
&&T_{\gamma N}= e^2\left\{ A_1 \vepsprime\cdot\veps 
             +A_2 \vepsprime\cdot\vkay \, \veps\cdot\vkayprime 
             +iA_3 \vsigma\cdot (\vepsprime\times\veps)
             +iA_4 \vsigma\cdot (\vkayprime\times\vkay)\, \vepsprime\cdot\veps
     \right. \nonumber \\
 & & \left.
\!\!\!\!\!\!\!\!\!  
     +iA_5 \vsigma\cdot [(\vepsprime\times\vkay)\, \veps\cdot\vkayprime
            -(\veps\times\vkayprime)\, \vepsprime\cdot\vkay] 
     +iA_6 \vsigma\cdot [(\vepsprime\times\vkayprime)\, \veps\cdot\vkayprime
                        -(\veps\times\vkay)\, \vepsprime\cdot\vkay] \right\}.
\label{eq:Ti}
\end{eqnarray}
The contribution at $\nu =-1$ ($O(Q^2)$) yields the low-energy theorem
for Compton scattering from a single nucleon---the Thomson limit:

\begin{equation}
T_{\gamma N}=-\veps \cdot \vepsprime \frac{{\cal Z}^2 e^2}{M},
\label{eq:OQ2}
\end{equation}
where ${\cal Z}$ is the nucleon charge.  The tree graphs discussed
under 1 and 2 for $\nu =0$ in Section~\ref{sec-kernelcounting} have
been known for a long time.  The amplitude including these two sorts
of contributions corresponds to a non-relativistic expansion of
nucleon Born terms in a relativistic model with derivative coupling of
the photon and Pauli magnetic moment interactions, plus a pion pole
diagram.  The differential cross section for scattering on the nucleon
that results from these diagrams in the $g_A=0$ limit is known as the
Powell cross section.  Adding diagrams of type 3 completes the
$O(Q^3)$ amplitude in $\cpt$ without an explicit delta field
\cite{bkmrev,ulf1,ulf2}.

Defining $\barf =\omega/m_\pi$ and $t=-2{\barf ^2}(1-\cos \theta)$,
where $\theta$ is the center-of-mass angle between the incoming and
outgoing photon momenta, one finds \cite{bkmrev,ulf1,ulf2}:

\begin{eqnarray}
A_1 &=& -\frac{{\cal Z}^2}{M}
        +\frac{g_A^2m_\pi}{8\pi f_\pi^2} 
         \left\{ 1- \sqrt{1-\barf^2}
                 +\frac{2-t}{\sqrt{-t}} 
                  \left[\frac{1}{2} \arctan \frac{\sqrt{-t}}{2}
                        -I_1(\barf,t) \right]\right\},
\nonumber \\
A_2 &=& \frac{{\cal Z}^2}{M^2\omega}
        -\frac{g_A^2}{8\pi f_\pi^2 m_\pi} 
         \frac{2-t}{(-t)^{3/2}} 
         \left[I_1(\barf,t)- I_2(\barf,t)\right],
\nonumber \\
A_3 &=& \frac{\omega}{2M^2} 
[{\cal Z}({\cal Z}+2\kappa )-({\cal Z}+\kappa)^2 \cos \theta]
        +\frac{{(2{\cal Z} -1)}g_Am_\pi}{8\pi^2 f_\pi^2} \frac{\barf t}{1-t}
\nonumber \\
    & &  +\frac{g_A^2m_\pi}{8\pi^2 f_\pi^2} 
         \left[ \frac{1}{\barf} \arcsin^2\barf- \barf +2\barf^4 
\sin^2 \theta I_3(\barf,t)\right],
\nonumber \\
A_4 &=& -\frac{({\cal Z}+\kappa )^2}{2M^2\omega}
        +\frac{g_A^2}{4\pi^2 f_\pi^2m_\pi} I_4(\barf,t),
\nonumber \\
A_5 &=& \frac{({\cal Z}+\kappa )^2}{2M^2\omega}
        -\frac{{(2{\cal Z} -1)}g_A}{8\pi^2 f_\pi^2 m_\pi} \frac{\barf}{(1-t)}
        -\frac{g_A^2}{8\pi^2 f_\pi^2m_\pi}
          [I_5(\barf,t)-2\barf^2\cos\theta I_3(\barf,t)],
\nonumber \\
A_6 &=& -\frac{{\cal Z}({\cal Z}+\kappa )}{2M^2\omega}
        +\frac{{(2{\cal Z} -1)}g_A}{8\pi^2 f_\pi^2 m_\pi}\frac{\barf}{(1-t)}
        +\frac{g_A^2}{8\pi^2 f_\pi^2m_\pi}
          [I_5(\barf,t)-2\barf^2 I_3(\barf,t)],
\label{eq:As}
\end{eqnarray} 
where
\begin{eqnarray}
I_1(\barf,t) &=& \int_0^1  dz \,
             \arctan \frac{(1-z)\sqrt{-t}}{2\sqrt{1-\barf^2 z^2}},
\nonumber \\
I_2(\barf,t) &=& \int_0^1  dz \,
             \frac{2(1-z)\sqrt{-t(1-\barf^2z^2)}}{4(1-\barf^2 z^2)-t(1-z)^2},
\nonumber \\
I_3(\barf,t) &=& \int_0^1  dx \, \int_0^1  dz \,
             \frac{x(1-x)z(1-z)^3}{S^3} 
             \left[ \arcsin \frac{\barf z}{R}+ \frac{\barf zS}{R^2}\right],
\nonumber \\
I_4(\barf,t) &=& \int_0^1  dx \, \int_0^1  dz \,
             \frac{z(1-z)}{S}\arcsin \frac{\barf z}{R},
\nonumber \\
I_5(\barf,t) &=& \int_0^1  dx \, \int_0^1  dz \,
             \frac{(1-z)^2}{S}\arcsin \frac{\barf z}{R},
\label{eq:Is}
\end{eqnarray}
with
\begin{equation}
S=\sqrt{1-\barf^2 z^2-t(1-z)^2x(1-x)}, \qquad R=\sqrt{1-t(1-z)^2x(1-x)}.
\end{equation}

Because there is no counterterm consistent with chiral symmetry and
gauge invariance at $O(Q^3)$, the sum of all loop graphs at this order
is finite.  Note that the loops contribute the same for protons and
neutrons. Note also that, partly because of these loops, the Compton
scattering amplitude is non-analytic in $\omega$ and in $\cos\theta$.
For $\omega$ sufficiently small a power-series expansion in $\omega$
can be made. Up to an overall factor of $\alpha_{\rm em}$ the
coefficient of the term proportional to $\omega^2$ in the invariant
amplitude $A_1$ is then $\alpha_N$, and that proportional to $\omega^2
\cos \theta$ is $\beta_N$. In this way we can extract the $\cpt$
predictions for the polarizabilities which were displayed in
Eq.~(\ref{eq:betaOQ3}) \cite{ulf1}. These effects of the pion cloud
dominate over short-range effects represented by higher-order contact
terms. The lack of isospin dependence in loops then implies that in
general the isoscalar polarizabilities are larger than the isovector
ones. Furthermore, a numerical factor of 10 decreases the magnetic
polarizabilities.  Thus, in general we have, $\alpha_p \simeq \alpha_n
\gg \beta_p\simeq \beta_n$.  The agreement with the experimental
values (\ref{protpolexpt}) and (\ref{neutpolexpt1}) is very good.

\begin{figure}[t,h,b,p]
  \vspace{0.3cm} \epsfysize=7cm \centerline{\epsffile{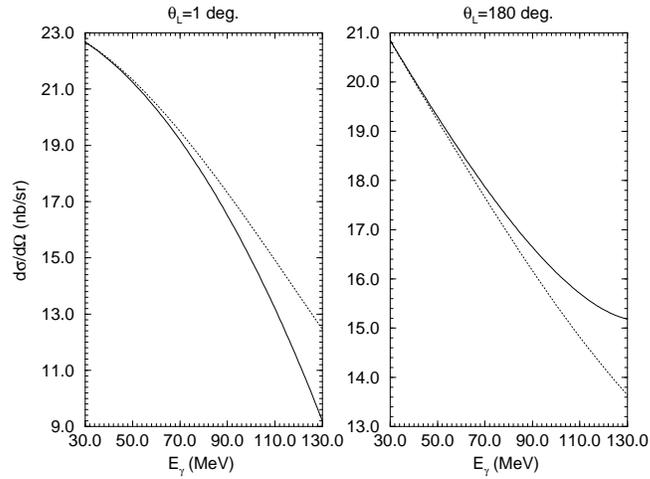}}
  \centerline{\parbox{11cm}{\caption{\label{fig-polapprox} Unpolarized
        cross section for Compton scattering on the proton in
        deltaless $\cpt$ to $O(Q^3)$ as a function of the photon
        energy at two laboratory angles, $\theta_L=1^o, 180^o$:
        polarizability approximation (dashed line) compared to full
        amplitude (solid line).}}}
\end{figure}

The difference between the full $O(Q^3)$ amplitude and the amplitude
truncated at the level of the polarizabilities increases with energy.
At $E_\gamma \sim 100$ MeV the typical error is 10\%.  However, the
polarizability approximation favors forward directions, since at
forward (backward) angles it overestimates (underestimates) the full
amplitude.  Two extreme cases are shown in Fig. \ref{fig-polapprox}.
Both the full amplitude and the polarizability approximation are in
relatively good agreement with data on the proton up to $\sim 120$ MeV
\cite{bkmrev,babusci}, although the agreement deteriorates with
increasing energy and angle \cite{babusci}.

\begin{figure}[t,h,b,p]
   \vspace{0.5cm}
   \epsfysize=5cm
   \centerline{\epsffile{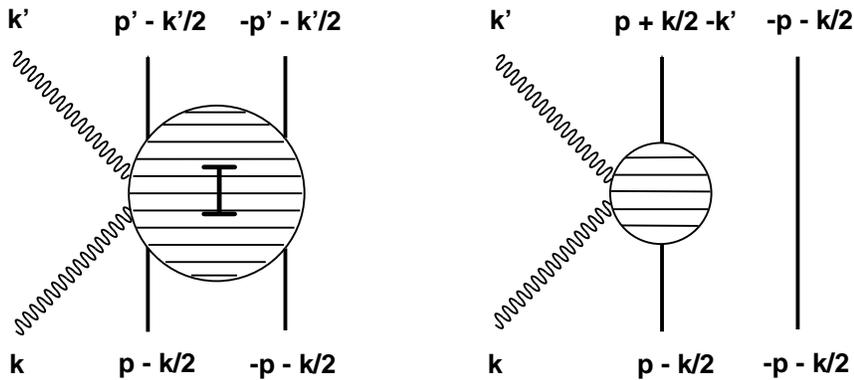}}
   \centerline{\parbox{11cm}{\caption{\label{figcaca}
Labelling of 3-momenta for 2- and 1-nucleon mechanisms in the
$\gamma NN$ c.m. frame. 
  }}}
\end{figure}

The only modification which must be made to the photon-nucleon
amplitude (\ref{eq:Ti}) for use in Compton scattering on the deuteron
is that it must be boosted from the $\gamma N$ c.m. frame to the
$\gamma NN$ c.m. frame. To the order we are working this is easily
accomplished, in one of two equivalent ways. Since only the $s$ and
$u$-channel nucleon-pole graphs depend on the nucleon momenta, these
graphs can be computed in a frame in which the total $\gamma N$
three-momentum is non-zero (see Fig. \ref{figcaca}), or the $O(Q^2)$
amplitude (\ref{eq:OQ2}) can be boosted by making the appropriate
substitutions to ensure that the polarization vectors $\veps$ and
$\vepsprime$ stay orthogonal to $\vkay$ and $\vkayprime$ in the new
frame~\cite{KW}. In both cases an additional term:

\begin{equation}
T_{boost}=-
\frac{{\cal Z}^2 e^2}{2{M^2}\omega}
\lbrack 
\veps\cdot\vkayprime \, \vepsprime\cdot\vkay +
{2}(\veps\cdot{\vec p} \, \vepsprime\cdot\vkay +
\veps\cdot\vkayprime \, \vepsprime\cdot{\vec p})
\rbrack
\end{equation}
must be added to the one-body amplitude (\ref{eq:Ti}) when we are 
computing Compton scattering from the two-nucleon system at $O(Q^3)$.

As for higher-order mechanisms, effects of the $O(Q^4)$ terms in
$\cpt$ on the nucleon polarizabilities have been studied.  In
Ref. \cite{ulf2} they were estimated to be considerable, especially in
the case of the neutron magnetic polarizability:

\begin{eqnarray}
\alpha_p=10.5 \pm 2.0 \times 10^{-4} \, {\rm fm}^3; &\qquad&
\alpha_n=13.4 \pm 1.5 \times 10^{-4} \, {\rm fm}^3; \label{eq:alphaOQ4}\\
\beta_p=3.5 \pm 3.6 \times 10^{-4} \, {\rm fm}^3;&\qquad&
\beta_n=7.8 \pm 3.6 \times 10^{-4} \, {\rm fm}^3. \label{eq:betaOQ4}
\end{eqnarray}
Note that the uncertainties here are {\it theoretical} and arise from
the use of resonance-saturation to estimate some of the $O(Q^4)$
$\cpt$ counterterms.  The largest uncertainty concerns the effects of
the delta isobar, which has a large electromagnetic transition to the
nucleon via an M1 operator, and therefore affects $\beta_N$
significantly. Consequently, the resonance-saturation
estimate~\cite{ulf2} of its tree-level contribution to a $\gamma\gamma
NN$ counterterm is large, but in the case of $\beta_p$ it is largely
canceled by the $O(Q^4)$ $\pi N$ loop contribution.

Attempts have been made to better determine the delta effects by
inclusion of an explicit $\Delta$
field~\cite{jenkinsetal,hemmert}. This amounts to treating
$M_\Delta-M$ as another quantity which is small relative to $\lsc$.
The $\cpt$ expansion is then an expansion in the ratio of $p$, $m_\pi$
{\it and} $M_\Delta-M$ to $\lsc$. These three small ratios are
generically labelled $\epsilon$.  This $\epsilon$ expansion is
certainly appropriate for the full amplitude at $p \sim m_\pi$, but
possibly not very efficient for low-energy quantities such as the
polarizabilities.  The result of the explicit inclusion of the delta
field is a shift of delta effects to lower orders. To $O(\epsilon^3)$
in this modified expansion there are large tree and loop contributions
from delta graphs to $\beta_N$, but now there are no cancelations
against $\pi N$ graphs.  There are also considerable loop
contributions to $\alpha_N$. The $O(\epsilon^3)$ result
is~\cite{hemmert}:
\begin{eqnarray}
\alpha_p=\alpha_n&=&16.4 \times 10^{-4} \, {\rm fm}^3;\label{eq:alphaOQ3del}\\ 
\beta_p=\beta_n&=& 9.1 \times 10^{-4} \, {\rm fm}^3. \label{eq:betaOQ3del}
\end{eqnarray}
One can only hope that an $O(\epsilon^4)$ calculation will bring these
values closer to the experimental ones of (\ref{protpolexpt}) and
(\ref{neutpolexpt1}). Thus far, calculations to $O(Q^4)$ in the
deltaless theory and to $O(\epsilon^3)$ in the deltaful theory have
been limited to polarizabilities and to forward amplitudes, and so, as
yet, in neither of these theories is there a calculation of the
invariant functions $A_1$--$A_6$ for arbitrary photon scattering
angle.

\subsection{Two-nucleon contributions at order $Q^3$ ($\nu = 0$)}

The power counting discussed in Section~\ref{sec-counting} shows that
when nuclear targets are involved a new contribution is present at the
same order ($\nu =0$) that the polarizabilities first appear: the
two-nucleon contributions depicted in Fig.~\ref{fig2}.  In principle,
then, we cannot hope to extract neutron polarizabilities without a
theory for the two-nucleon contributions.  Moreover, comparison of
diagrams in Figs. \ref{fig1}(e) and \ref{fig2} show that the same
physics that produces dominant contributions to the polarizabilities
generates the dominant two-nucleon contributions. The only difference
lies in the destination of the pion emitted by one nucleon: if it is
absorbed on the same nucleon it contributes to the polarizability; if
it lands on another nucleon, it generates a two-nucleon contribution.
In this section we compute the dominant two-nucleon contributions in
$\cpt$---they involve no new parameters. It is also important to
notice that these $O(Q^3)$ two-nucleon Feynman diagrams are the same
as those at $O(\epsilon^3)$ in a theory in which the $\Delta$ is
included explicitly.  Only at next order would we find differences.

The two-nucleon amplitude from the diagrams in Fig. \ref{fig2} is

\begin{equation}
T_{\gamma NN}^{2N}\;=\;-\frac{{e^2}{\ga ^2}}{2 f_\pi^2}
\; ({\vec\tau}^{\; 1} \cdot{\vec\tau}^{\; 2}-\tau^{1}_{z}\tau^{2}_{z})
\; ( t^{(a)}+
t^{(b)}+
t^{(c)}+
t^{(d)}+
t^{(e)})
\label{total}
\end{equation}

\begin{eqnarray}
{t^{(a)}}&= &
\frac{{\veps\cdot\vsigone}\;{\vepsprime\cdot\vsigtwo}}
{2\lbrack {\omega^2}-{m_\pi^2}-
(\vpee -\vpeeprime +{\frac{1}{2}(\vkay +\vkayprime )})^2 \rbrack}
+ (1\;\leftrightarrow\; 2) 
\label{ta}\\
{t^{(b)}}&= &  
\frac{{\veps\cdot\vepsprime}\;
\vsigone\cdot ( \vpee -\vpeeprime -{\frac{1}{2}(\vkay -\vkayprime )} )
        \vsigtwo\cdot ( \vpee -\vpeeprime +{\frac{1}{2}(\vkay -\vkayprime )})}
{2\lbrack (\vpee -\vpeeprime -{\frac{1}{2}(\vkay -\vkayprime )})^2 
+{m_\pi^2} \rbrack
\lbrack (\vpee -\vpeeprime +{\frac{1}{2}(\vkay -\vkayprime )})^2  
+{m_\pi^2} \rbrack}
+ (1\;\leftrightarrow\; 2) 
\label{tb}\\
{t^{(c)}}&= &  
-\frac{{\vepsprime\cdot (\vpee -\vpeeprime +{\frac{1}{2}\vkay})}\;
          \vsigone\cdot\veps\;
 \vsigtwo\cdot ( \vpee -\vpeeprime +{\frac{1}{2}(\vkay -\vkayprime )} )}
{\lbrack {\omega^2}-{m_\pi^2}- (\vpee -\vpeeprime 
+{\frac{1}{2}(\vkay +\vkayprime )})^2 \rbrack
\lbrack (\vpee -\vpeeprime +{\frac{1}{2}(\vkay -\vkayprime )})^2  
+{m_\pi^2} \rbrack}
+ (1\;\leftrightarrow\; 2) 
\label{tc}\\
{t^{(d)}}&= & 
-\frac{{\veps\cdot (\vpee -\vpeeprime +{\frac{1}{2}\vkayprime} )}\;
\vsigone\cdot ( \vpee -\vpeeprime -{\frac{1}{2}(\vkay -\vkayprime )} )
              \vsigtwo\cdot\vepsprime }
{\lbrack {\omega^2}-{m_\pi^2}- (\vpee -\vpeeprime +
{\frac{1}{2}(\vkay +\vkayprime )})^2 \rbrack
\lbrack (\vpee -\vpeeprime -{\frac{1}{2}(\vkay -\vkayprime )})^2  
+{m_\pi^2} \rbrack}+ (1\;\leftrightarrow\; 2) 
\label{td}\\
{t^{(e)}}&= &  
\frac{2
{\veps\cdot (\vpee -\vpeeprime +{\frac{1}{2}\vkayprime})\;
\vepsprime\cdot (\vpee -\vpeeprime +{\frac{1}{2}\vkay})}\;
\vsigone\cdot ( \vpee -\vpeeprime -{\frac{1}{2}(\vkay -\vkayprime )})\; 
\vsigtwo\cdot ( \vpee -\vpeeprime +{\frac{1}{2}(\vkay -\vkayprime )} )}
{\lbrack {\omega^2}-{M_\pi^2}- (\vpee -\vpeeprime +
{\frac{1}{2}(\vkay +\vkayprime )})^2 \rbrack
\lbrack (\vpee -\vpeeprime -{\frac{1}{2}(\vkay -\vkayprime )})^2  
+{m_\pi^2} \rbrack
\lbrack (\vpee -\vpeeprime +{\frac{1}{2}(\vkay -\vkayprime )})^2  
+{m_\pi^2} \rbrack}\nonumber \\
&+& (1\;\leftrightarrow\; 2) 
\label{te}
\end{eqnarray}

We calculate the cross section for Compton scattering on the deuteron
including the single-scattering and two-nucleon mechanisms described
above.  Our calculation represents therefore the full $\nu =0$, or
order $Q^3$, $\cpt$ {\it predictions} for Compton scattering on the
deuteron.

\section{Results and discussion}
\label{sec-results}

To calculate the amplitude for Compton scattering on the deuteron
we must take the residue of Eq.~(\ref{eq:ComptonGF}) at the initial
and final bound-state pole. This leads to the amplitude

\begin{equation}
T^{\gamma d}=\langle \psi_d| K_{\gamma \gamma} |\psi_d \rangle,
\end{equation}
where $|\psi_d \rangle$ is the deuteron wave function and we may
eliminate the isospin factor by making use of the relation $\langle
\psi_d| ({\vec\tau}^{\; 1} \cdot{\vec\tau}^{\;
  2}-\tau^{1}_{z}\tau^{2}_{z}) |\psi_d \rangle =-2$.

In Section~\ref{sec-calc} we wrote down both the one and two-body
contributions to $K_{\gamma \gamma}$ at $O(Q^3)$ in the chiral
expansion. If we make the same separation here then we can write down
the following integrals for $T_{\gamma d}$ in the $\gamma d$
center-of-mass frame:

\begin{eqnarray}
T^{\gamma d}_{M' \lambda' M \lambda}(\vkayprime,\vkay)&=& \int
\frac{d^3p}{(2 \pi)^3} \, \, \psi_{M'}\left( \vpee + \frac{\vkay -
\vkayprime}{2}\right) \, \, T^{\gamma d \, \, c.m.}_{\gamma N_{\lambda'
\lambda}}(\vkayprime,\vkay) \, \, \psi_M(\vpee)\nonumber\\ 
&+& \int \frac{d^3p \, \, d^3p'}{(2 \pi)^6} \, \, \psi_{M'}(\vpeeprime) \, \,
T^{2N}_{\gamma NN_{\lambda' \lambda}}(\vkayprime,\vkay) \, \, \psi_M(\vpee)
\label{eq:gammad}
\end{eqnarray}
where $M$ ($M'$) is the initial (final) deuteron spin state, and
$\lambda$ ($\lambda'$) is the initial (final) photon polarization
state, and $\vkay$ ($\vkayprime$) the initial (final) photon
three-momentum, which are constrained to
$|\vkay|=|\vkayprime|=\omega$. Note that the single-scattering $\gamma
N$ $T$-matrix $T^{\gamma d \, \, c.m.}_{\gamma N}$ must be evaluated
in the $\gamma d$ center-of-mass frame. In $\cpt$ this is
straightforward, as discussed in the previous section. Note also that
all sums over the spins of the two nucleons have been suppressed here.

In general, for the wave function $\psi$ we use the energy-independent
Bonn OBEPQ wave function parameterization which is found in
Ref.~\cite{Bonnrep}.  The photon-deuteron $T$-matrix (\ref{eq:gammad})
is then calculated and the laboratory differential cross section
evaluated directly from it:

\begin{equation}
  \frac{d \sigma}{d \Omega_L}=\frac{1}{16 \pi^2}
  \left(\frac{E_\gamma'}{E_\gamma}\right)^2 \frac{1}{6} \sum_{M'
    \lambda' M \lambda} |T^{\gamma d}_{M' \lambda' M \lambda}|^2,
\end{equation}
where $E_\gamma$ is the initial photon energy in the laboratory frame,
and is related to $\omega$, the photon energy in the $\gamma d$
center-of-mass frame, via:

\begin{equation}
\omega=\frac{E_\gamma}{\sqrt{1 + 2 E_\gamma/M_d}};
\end{equation}
and $E_\gamma'$ is the final photon energy in the laboratory frame:

\begin{equation}
E_\gamma'=\frac{E_\gamma M_d}{M_d + E_\gamma (1 - \cos \theta_L)}.
\end{equation}
Convergence tests indicate that with the numbers of quadratures chosen
the cross section evaluated in this fashion is numerically accurate at
about the 1\% level. Of course, this error does not include the
theoretical error from uncertainties due to different deuteron wave
functions, and the effect of higher-order terms in the $\cpt$
calculation of the kernel. These errors will be discussed further
below.

The rest of this section is divided into four subsections.  First we
display our results for this straightforward $O(Q^3)$ calculation.
Results are presented at laboratory photon energies of 49, 69, and 95
MeV. At the lower two energies we compare with the experimental data
of Lucas~\cite{lucas}. Data at $E_\gamma=95$ MeV are expected soon
from a recent Saskatoon tagged photon Compton scattering
experiment~\cite{SAL}. Second, we observe that at lower energies
contributions due to $NN$ states with relative momenta of order
$\sqrt{M \omega}$ lead to certain diagrams being enhanced over the
value expected from the power counting of Section~\ref{sec-bcpt}.
Indeed, as discussed in Section~\ref{sec-counting}, we expect this
power counting to break down for $\omega \sim m_\pi^2/M$. We estimate
the effect of these contributions by calculating the impact some of
them have on the differential cross section for photon-deuteron
scattering.  Third, we display the sensitivity of our results to the
choice of deuteron wave function. Re-calculating our amplitudes with
$\psi_d$ defined by solving the Schr\"odinger equation with the Nijm93
potential \cite{Nijm93} allows us to estimate the error associated
with uncertainties in the $NN$ interaction. Finally, in order to see
the size of the effects we can expect in an $O(Q^4)$ calculation of
this process, we arbitrarily include two terms which contribute to the
$O(Q^4)$ calculation: counterterms which shift the values of the
nucleon polarizabilities from the $O(Q^3)$ result.

\subsection{Results at $O(Q^3)$}

\begin{figure}[t,h,b,p]
   \vspace{0.5cm} \epsfysize=7.5cm
   \centerline{\epsffile{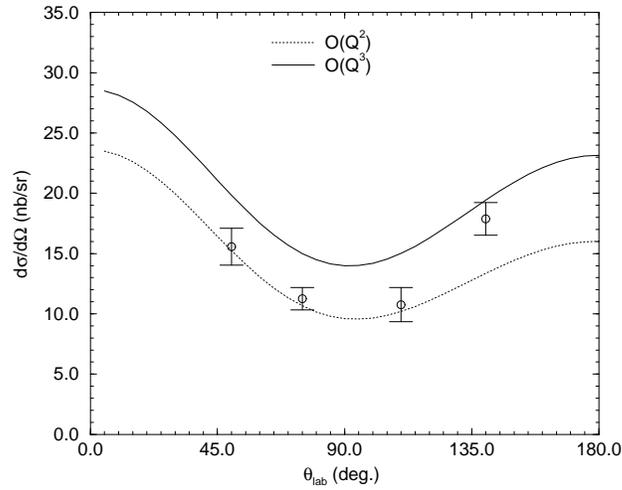}}
   \centerline{\parbox{11cm}{\caption{\label{fig6} Results of the
   $O(Q^2)$ (dotted line) and $O(Q^3)$ (solid line) calculations
   at a photon laboratory energy of 49 MeV. The data points of
   Ref.~\protect{\cite{lucas}} are also shown.}}}
\end{figure}

\begin{figure}[t,h,b,p]
   \vspace{0.5cm} \epsfysize=7.5cm
   \centerline{\epsffile{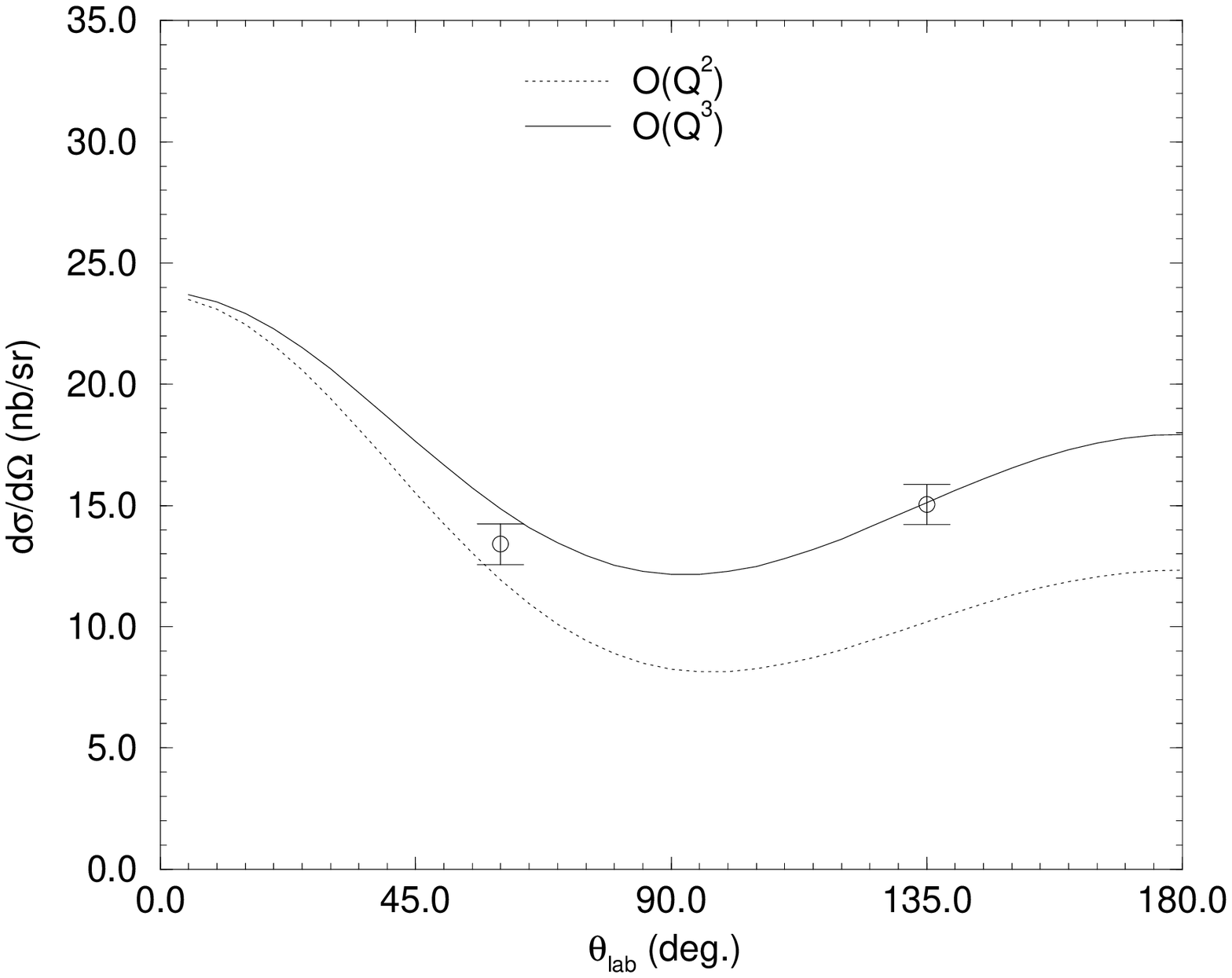}}
   \centerline{\parbox{11cm}{\caption{\label{fig7} Results of the
   $O(Q^2)$ (dotted line) and $O(Q^3)$ (solid line) calculations
   at a photon laboratory energy of 69 MeV. The data points of
   Ref.~\protect{\cite{lucas}} are also shown.}}}
\end{figure}

\begin{figure}[t,h,b,p]
   \vspace{0.5cm} \epsfysize=7.5cm
   \centerline{\epsffile{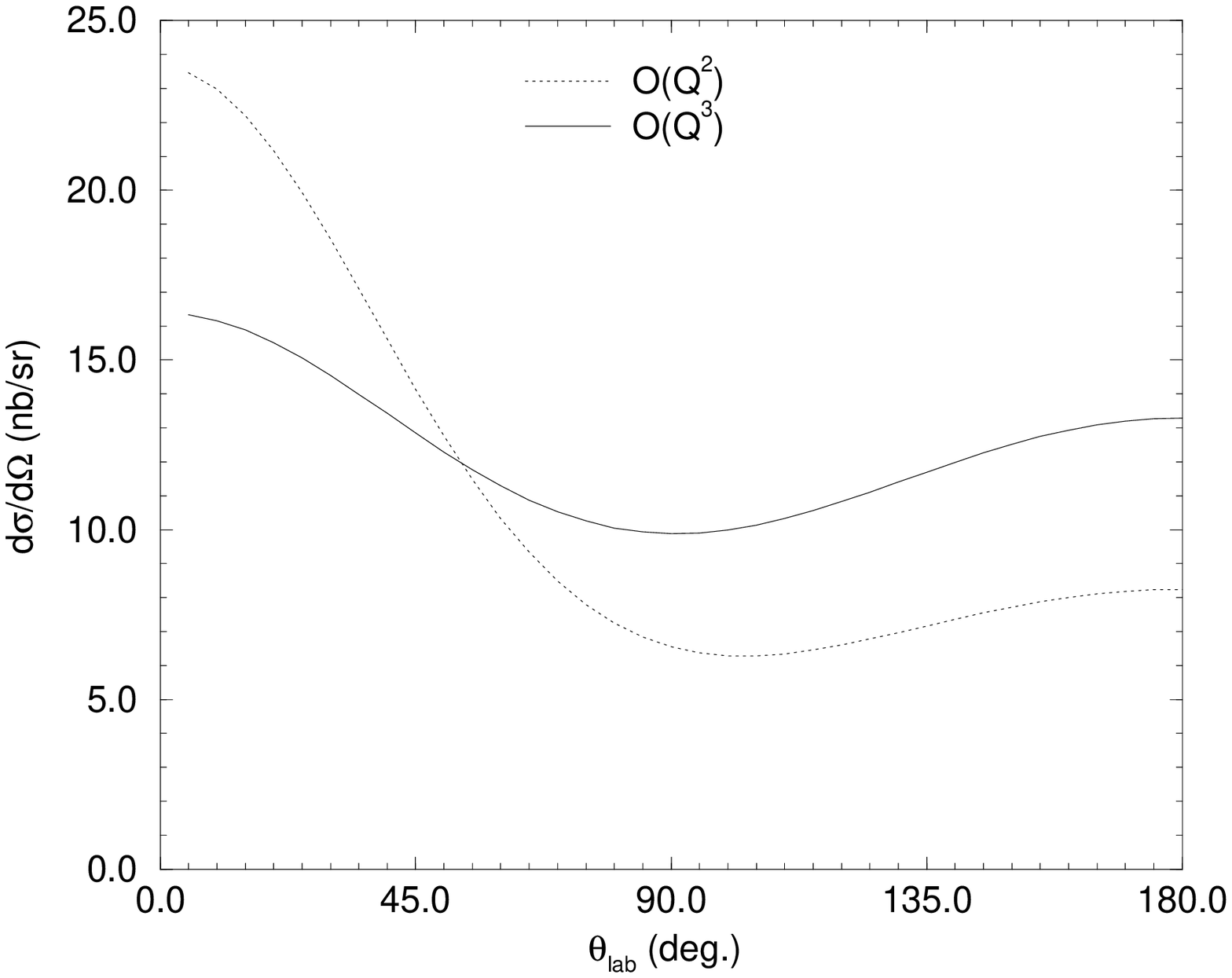}}
   \centerline{\parbox{11cm}{\caption{\label{fig8} Results of the
   $O(Q^2)$ (dotted line) and $O(Q^3)$ (solid line) calculations
   at a photon laboratory energy of 95 MeV.}}}
\end{figure}

In figures~\ref{fig6}, \ref{fig7} and \ref{fig8} below we display our
results at 49, 69, and 95 MeV. For comparison we have included the
calculation at $O(Q^2)$, where the second contribution in
Eq.~(\ref{eq:gammad}) is zero, and the $\gamma N$ $T$-matrix in the
single-scattering contribution is given by the Thomson term on a
single nucleon, Eq.~(\ref{eq:OQ2}).

The curves show that the correction from the $O(Q^3)$ terms gets
larger as $\omega$ is increased, as was to be expected. Indeed, 
while at lower energies corrections are relatively small, in
the 95 MeV results the correction to the differential cross section
from the $O(Q^3)$ terms is of order 50\%, although the
contribution of these terms to the {\it amplitude} is of roughly the
size one would expect from the power counting: about 25\%.
Nevertheless, it is clear, even from these results, that
this calculation must be performed to $O(Q^4)$ before
conclusions can be drawn about 
polarizabilities from data
at photon energies of order $m_\pi$. This is in accord with
similar convergence properties for the analogous calculation
for threshold pion photoproduction on the deuteron~\cite{silas2}.

We have also shown the six Illinois data points at 49 and 69
MeV~\cite{lucas}. Statistical and systematic errors have been added in
quadrature. It is quite remarkable how well the $O(Q^2)$ calculation
reproduces the 49 MeV data. However, it is clear that the agreement at
forward angles is somewhat fortuitous, as there are significant
$O(Q^3)$ corrections.  Meanwhile, the agreement of the $O(Q^3)$
calculation with the 69 MeV data is very good, although only limited
conclusions can be drawn, given that there are only two data points,
each with sizeable error bars.

The energy-dependence of our results at a given angle is due to a
combination of two mechanisms. First, at backward angles the
$O(Q^2)$ deuteron Compton scattering amplitude has 
energy-dependence, due to it is proportionality to the non-relativistic
deuteron charge form factor at a momentum transfer
\begin{equation}
\vec{q}=\vkay - \vkayprime.
\end{equation}
At forward angles this energy-dependence is minimal, since $|\vec{q}\,
|^{\; 2}$ is proportional to $\omega^2 (1 - \cos \theta)$. Second, at
both forward and backward angles the amplitude also has significant
energy-dependence due to the structure of the single-nucleon amplitude
(\ref{eq:Ti}) as a function of $\omega$.  This is particularly
apparent at $\theta_{cm}=0$, where the $O(Q^2)$ amplitude is, in fact,
energy-independent, and we can see that the polarizabilities play a
significant role in suppressing the forward differential cross section
as the photon energy increases.

\begin{figure}[t,h,b,p]
   \vspace{0.5cm} \epsfysize=7.5cm
   \centerline{\epsffile{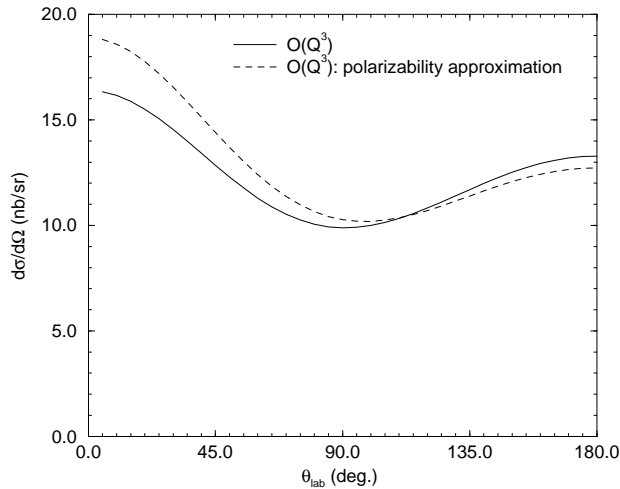}}
   \centerline{\parbox{11cm}{\caption{\label{fig-deutpolapprox}
   Unpolarized cross section for Compton scattering on the deuteron in
   deltaless $\cpt$ to $O(Q^3)$ at 95 MeV.  The polarizability
   approximation to the single-nucleon amplitude (dashed line) is to
   be compared with the full $O(Q^3)$ calculation (solid line).}}}
\end{figure}

Our results are qualitatively not very different from other existing
calculations.  At 49 and 69 MeV our $O(Q^3)$ results are very close to
those in Ref. \cite{wilbois} and a few nb/sr higher, especially at
back angles, than those of Refs. \cite{levchuk,jerry} (which are
similar at these energies).  At 95 MeV our $O(Q^3)$ result is close to
that of Ref. \cite{levchuk}, higher by several nb/sr at back angles
than Ref. \cite{jerry}, and several nb/sr lower than the calculation
with no polarizabilities of Ref. \cite{wilbois}~\footnote{At this
energy Ref. \cite{wilbois} only presents results with
$\alpha_p+\alpha_n=\beta_p+\beta_n=0$, which in turn are considerably
less forward peaked than the corresponding calculation of
Ref. \cite{levchuk}.}.  Comparing to the calculations of deuteron
Compton scattering in the KSW formulation of effective field
theory~\cite{martinetal}, we see that the result of Ref.
\cite{martinetal} is significantly lower than those presented here at
both 49 and 69 MeV. At 49 MeV the agreement of
Ref. \cite{martinetal}'s calculation with the data is better than
ours. We shall show in the next section that this is partly because 49
MeV is at the lower end of the domain of applicability of the Weinberg
formulation.  At 69 MeV our calculation does a slightly better job of
reproducing the (two) data points available. The qualitative agreement
among these calculations is a reflection of the similarities of
mechanisms involved.  Ours is however the only calculation to
incorporate the full single-nucleon amplitude instead of its
polarizability approximation.  As shown in
Figure~\ref{fig-deutpolapprox} our tendency to higher cross sections
in the backward directions is at least in part due to this feature, as
expected from the discussion in Sect. \ref{sec-calc}.

\subsection{The breakdown of the power counting at low energies}

In this section we examine the breakdown of the power counting defined
by Eq.~(\ref{nu}) at low energies. Recall that this power counting was
only derived for photon energies of order $m_\pi$. Here we will
calculate one portion of a particular diagram, and show how the
contribution of that diagram becomes enhanced beyond the expectations
of power counting if the photon energy becomes too low.

Consider the diagram shown in Figure~\ref{fig-s-channelpole}.  If the
photon energy $\omega$ is of order $m_\pi$ and therefore much larger
than the nuclear binding and the nucleon kinetic energies then the
denominator of the intermediate-state $NN$ propagator can be
approximated by $1/\omega$. Effects due to the motion of bound
nucleons then appear as higher-order vertices in the $\cpt$
Lagrangian, while effects due to binding also occur at higher order in
the calculation. This is the approach adopted in this work. However,
as $\omega$ is decreased, replacing the denominator in the two-nucleon
propagator by $\omega$ is no longer a valid approximation. It becomes
necessary to retain the full energy-dependence for dynamical nucleons,
rather than use the heavy-baryon propagator which treats the nucleon
as a static source. Thus, in this section we evaluate this graph using
a propagator which retains the full momentum dependence for the
two-nucleon intermediate state. In heavy-baryon chiral perturbation
theory this means replacing the standard heavy baryon $\cpt$
propagator for two nucleons with energy $\omega$ by one in which a set
of nominally higher-order corrections representing the motion of the
nucleons has been resummed~\cite{weinnp}:
\begin{equation} 
\frac{1}{\omega} \longrightarrow \frac{M}{M \omega - {\vec p}^{\; 2}},
\end{equation}
with $\vpee$ the relative three-momentum of the two-nucleon state. 

To simplify matters we consider only the vertices from the $\chi PT$
Lagrangian ${\cal L}^{(1)}$ which arise from electric coupling of the
photon to the nucleon. This gives:

\begin{figure}[t,h,b,p]
   \vspace{0.5cm} \epsfysize=5.5cm
   \centerline{\epsffile{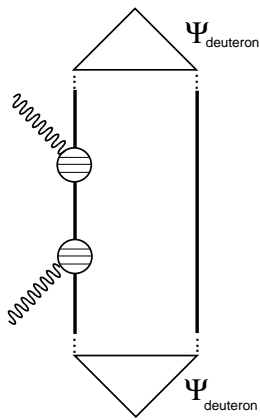}}
   \centerline{\parbox{11cm}{\caption{\label{fig-s-channelpole}
   Contribution to Compton scattering on the deuteron from the {\it
   s}-channel pole in photon-nucleon scattering.}}}
\end{figure}

\begin{eqnarray}
  &&T_{s-channel \, \, pole}=\nonumber\\
&&\quad \int \frac{d^3p}{(2 \pi)^3} \, \, \psi\left(\vpee +
\frac{\vkay - \vkayprime}{2} \right) \, \, \frac{\vepsprime \cdot (2 \vpee
+ \vkay - \vkayprime)}{M} \, \, \frac{M}{M (\omega - B) - \left(\vpee +
\frac{\vkay}{2}\right)^2 + i \eta} \, \, \frac{\veps \cdot (2 \vpee)}{M} \, \,
\psi(\vpee)\nonumber\\
\label{eq:NNoncont}
\end{eqnarray}
To estimate the size of effects arising from the dynamical nature
of the nucleons in the intermediate $NN$ state we have rewritten
the two-nucleon propagator as:

\begin{equation}
\frac{M}{M(\omega - B) - \left(\vpee + \frac{\vkay}{2}\right)^2 + i \eta}=P
\frac{M}{M(\omega - B) -\left(\vpee + \frac{\vkay}{2}\right)^2} - i M \pi
\delta \left(M \omega - M B -\left(\vpee + \frac{\vkay}{2}\right)^2 \right).
\end{equation}

For the case of low-energy photon-deuteron scattering the leading
contribution to the delta-function piece will be
$\delta(M \omega - {\vec p}^{\; 2})$. Thus, we may evaluate the leading part
of the delta-function piece of $T_{s-channel \, \, pole}$
as a two-dimensional integral over the solid angle $\Omega_p$:

\begin{equation}
T_{s-channel \, \, pole}=
\frac{i \pi}{2 \sqrt{M \omega} M}
\int \frac{d^3p}{(2 \pi)^3} \, \, \psi\left (\vpee +
\frac{\vkay - \vkayprime}{2} \right) \, \, 2 \vepsprime \cdot \vpee
\, \, \delta (|{\vec p}\; | - \sqrt{M \omega}) \, \,
2 \veps \cdot \vpee \, \, \psi (\vpee).
\label{eq:oq2.5piece}
\end{equation}

Naively this expression contributes to $T^{\gamma d}$ at an order
$\sqrt{\omega/\lsc}$ relative to the leading contribution.  However,
note that in fact this part of $T_{s-channel \, \, pole}$ is
suppressed beyond the level indicated by this straightforward
counting, since factors of the deuteron wave function at momenta of
order $\sqrt{M \omega}$ enter. Ultimately if $\omega \sim m_\pi$ these
factors drastically reduce this piece of $T_{s-channel \, \, pole}$,
rendering it quite small. However, at $\omega=49$ MeV, contributions
such as (\ref{eq:oq2.5piece}) can be sizeable~\footnote{The role of
  such contributions from on-shell intermediate-state nucleons in
  neutral pion photoproduction on the deuteron was emphasized by
  Wilhelm~\cite{willhelm}. In contrast to the case of photoproduction,
  in Compton scattering the incoming photon can be arbitrarily soft,
  so such contributions are unarguably important if the photon energy
  is low enough.}.

A contribution to Compton scattering on the deuteron similar to that
of (\ref{eq:oq2.5piece}) exists for the $u$-channel pole graph for
photon-nucleon scattering. The contribution of that graph to the
amplitude is, in fact, real, and the loop momentum that contributes to
that process is, at lowest order, $|{\vec p}\; |=i \sqrt{M
\omega}$. In fact, as explained in Refs.~\cite{martinetal} it is the
contribution from these momenta in the $u$-channel pole graph that
facilitates the (approximate) recovery of the Thomson limit for
low-energy photon scattering off the deuteron.

\begin{figure}[t,h,b,p]
  \vspace{0.3cm} \epsfysize=9.5cm \centerline{\epsffile{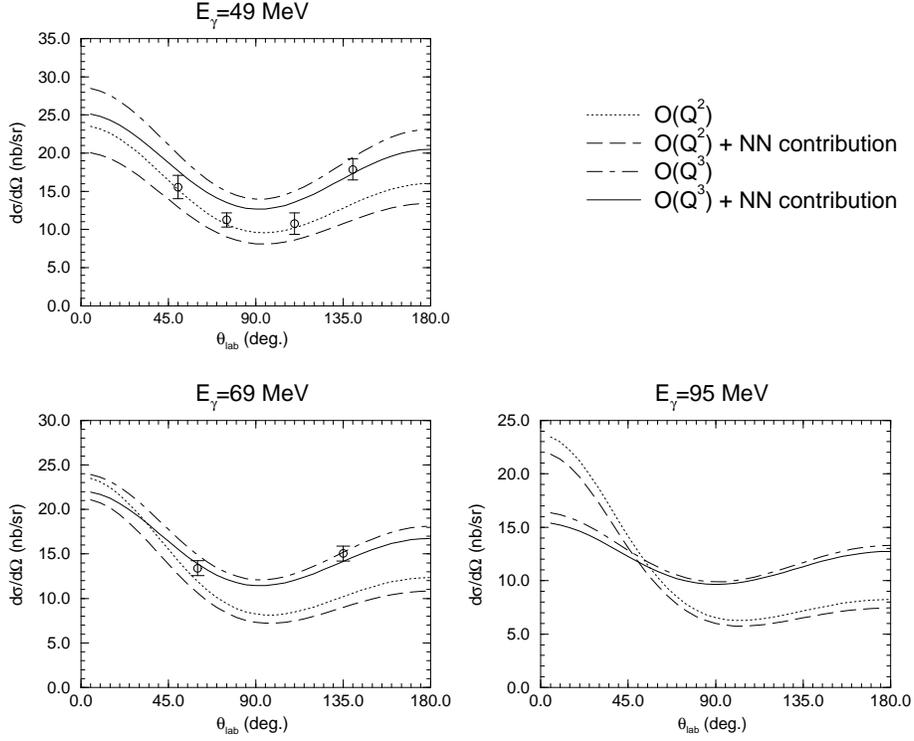}}
  \centerline{\parbox{11cm}{\caption{\label{fig-NNadded} Results of
        calculations at 49, 69, and 95 MeV both with and without the
        contribution from $NN$ intermediate states whose relative
        momenta are of order $\sqrt{M \omega}$. In each case the
        dotted line is the $O(Q^2)$ calculation, and the dot-dashed
        line is the $O(Q^3)$ calculation. Meanwhile the dashed
        line is an $O(Q^2)$ calculation with the contributions from
        $NN$ intermediate-states with momenta of $O(\protect{\sqrt{M
            \omega}})$ included, and the solid line is an $O(Q^3)$
        calculation with the same contributions included.
        Experimental data from Ref. \protect{\cite{lucas}} are also
        shown.}}}
\end{figure}

When these two contributions are added to the amplitude $T^{\gamma d}$
calculated above we obtain the results shown in
Figure~\ref{fig-NNadded}. There we have shown results for $O(Q^2)$ and
$O(Q^3)$, both with and without these mechanisms, at laboratory photon
energies of 49, 69, and 95 MeV. Clearly the contribution from these
two-nucleon states with relative momenta of order $\sqrt{M \omega}$
plays a significant role if $E_\gamma=49$ MeV. Since these effects are
higher-order in our power counting this apparently implies a breakdown
of our approach at $\omega \sim 50$ MeV. However, we observe that once
this $NN$ intermediate-state contribution is added the agreement with
the experimental data at 49 MeV is really quite good. This suggests
that it may be possible to devise an alternative power counting for
this regime, where an accounting is made not only of the way the
scales $\omega$, $p$ and $m_\pi$ appear in the amplitude, but also of
the role played by the additional scale $\sqrt{M \omega}$.

However, this low-energy regime is not a primary concern in our work
here.  We have instead focused on $\omega \sim m_\pi$, since it is
there that the best opportunity to extract the neutron polarizability
seems to lie. Our results show that as the photon energy is increased
the contribution from these on-shell $NN$ intermediate states becomes
significantly less, because of the suppression from the deuteron wave
function discussed above.  By the time $\omega=95$ MeV is reached this
higher-order effect is small: in fact, as we shall see below, it is
significantly smaller than other higher-order effects which could be
included.

\subsection{Sensitivity to choice of deuteron wave function}

\begin{figure}[t,h,b,p]
  \vspace{0.5cm} \epsfysize=8cm \centerline{\epsffile{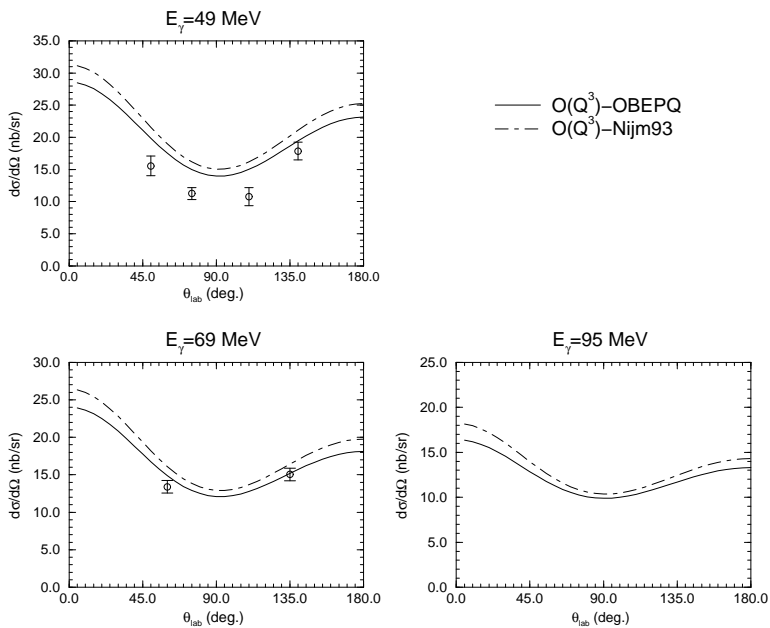}}
  \centerline{\parbox{11cm}{\caption{\label{fig-potcomp} Results of
        calculations at 49, 69, and 95 MeV using two different $NN$
        wave functions in the $O(Q^3)$ calculation. The solid line is
        the result using the OBEPQ wave function, and the dot-dashed
        line using the Nijm93 wave function.  Experimental data from
        Ref. \protect{\cite{lucas}} are also shown.}}}
\end{figure}

We have also performed the calculation of the photon-deuteron
differential cross section with a wave function found by solving
the Schr\"odinger equation with the Nijm93
potential~\cite{Nijm93}. A comparison of $O(Q^3)$ calculations
with the two different potentials at 49, 69, and 95 MeV is shown
in Figure~\ref{fig-potcomp}.  The results with the Nijm93 wave
function are in general a little higher than those found with
OBEPQ. However, the cross sections only differ at the 10\%
level. 

Further investigation shows that much of this discrepancy is due to
inconsistency between the physics of pion range in the $NN$ potentials
employed and the two-nucleon Compton scattering mechanisms involving
pions which we have calculated. In computing these graphs we employed
the $\pi NN$ vertex from the lowest-order chiral Lagrangian and the
axial coupling constant $g_A=1.26$. Therefore in essence we have used
a $\pi NN$ coupling in these graphs given by the Goldberger-Treiman
relation:
\begin{equation}
\left. f^2_{\pi NN} \right|_{\rm GT}=\frac{g_A^2 m_\pi^2}{16 \pi f_\pi^2}=
0.071.
\label{eq:GT}
\end{equation}

In contrast, the values of $f^2$ employed in the two $NN$ potentials
from which we obtain our deuteron wave functions are:
\begin{equation}
\left. f^2_{\pi NN} \right|_{\rm Bonn}=0.079; \quad \quad
\left. f^2_{\pi NN} \right|_{\rm Nijm93}=0.075.
\label{eq:fpinn}
\end{equation}
Effects beyond the leading-order result for $f_{\pi NN}$,
Eq.~(\ref{eq:GT}), will enter at orders beyond those we have
considered here and presumably move our $f_{\pi NN}$ towards the values
(\ref{eq:fpinn}).  Although such changes in $f_{\pi NN}$ would be
higher order in the chiral expansion, we note that in certain
circumstances they may enter raised to the fourth power in the
photon-deuteron cross section, and so might play a non-negligible
role.

A fully consistent resolution of this issue must await the use of an
accurate $\chi$PT deuteron wave function in the calculation of
deuteron Compton scattering. To address the issue in our current
calculation we multiplied the strength of our two-nucleon mechanisms
by two different factors in the calculations employing the two
different deuteron wave functions.  In this way the two-nucleon
current is made consistent with the value of the $\pi NN$ coupling
employed in the Nijm93 and OBEPQ potentials. We stress that the two
potentials still do not have the same strength at pion range, but at
least once such a scaling is employed the $NN$ mechanisms for Compton
scattering are always consistent with the one-pion exchange used to
obtain the deuteron wave function. This partial remediation of the
inconsistency reduced the wave function sensitivity to about 5\%
everywhere. This suggests to us that the difference between the
results obtained with the two different wave functions does not
represent a true sensitivity to short-distance dynamics, but rather a
need to treat the dynamics of pion range in the same way in all pieces
of the calculation.

As we shall see, this wave-function dependence is smaller than other
effects which are higher-order in the chiral expansion of the kernel.
Therefore at this stage we are not concerned about it.  Ultimately we
would like to perform this calculation of Compton scattering on the
deuteron with a deuteron wave function calculated consistently from an
$NN$ potential derived within the $\cpt$ framework~\cite{ray}.

\subsection{Effects of higher-order terms}

Finally, in order to test the sensitivity of our calculation to
higher-order effects we added a small piece of the $O(Q^4)$ amplitude
for Compton scattering off a single nucleon.  Specifically, we
modified the invariant functions $A_1$ and $A_2$ of Eq.~(\ref{eq:As})
to include additional terms.  The additional terms are

\begin{eqnarray}
\Delta A_1&=&(\Delta \alpha + \Delta \beta \cos \theta) \, \omega^2,
 \label{eq:DeltaA1}\\
\Delta A_2&=&-\Delta \beta. \label{eq:DeltaA2}
\end{eqnarray}
We emphasize that these are only two of a number of new terms
which will appear in the single-nucleon scattering amplitude
$T_{\gamma N}$ at $O(Q^4)$. Furthermore, a number of additional
two-body mechanisms must be included in $T_{\gamma NN}^{2N}$ in any
$O(Q^4)$ calculation of Compton scattering on the
deuteron. Nevertheless, here we calculate the differential
cross section with the terms (\ref{eq:DeltaA1}) and
(\ref{eq:DeltaA2}) in order to get a feel for the sensitivity of
our result to the presence of such higher-order terms.

The addition of (\ref{eq:DeltaA1}) and (\ref{eq:DeltaA2}) change
the polarizabilities in the calculation to 
\begin{equation}
\alpha=\alpha^{(Q^3)} + \Delta \alpha, \qquad \qquad
\beta=\beta^{(Q^3)} + \Delta \beta,
\label{mockpol}
\end{equation}
where $\alpha^{(Q^3)}$ and $\beta^{(Q^3)}$ are the $O(Q^3)$ values of
Eq.~(\ref{eq:betaOQ3}).

\begin{figure}[t,h,b,p]
   \vspace{0.5cm} \epsfysize=7.5cm
   \centerline{\epsffile{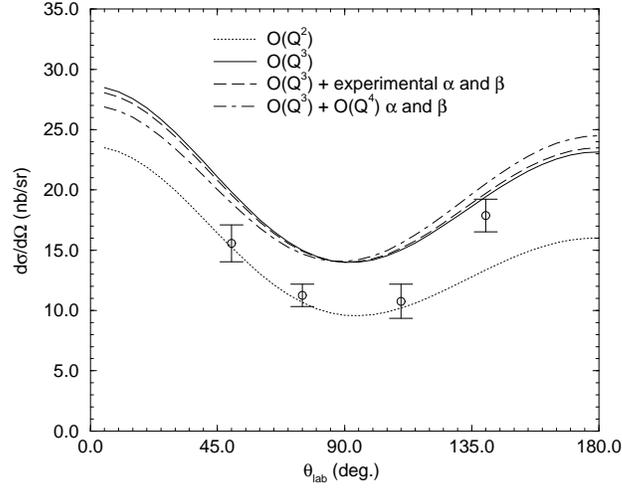}}
   \centerline{\parbox{11cm}{\caption{\label{fig-polcomp49} Results
   of calculations at 49 MeV using different values for the
   nucleon electromagnetic polarizabilities. The solid line is
   the result using the $O(Q^3)$ $\cpt$ value, the long-dashed
   line is the result using ``experimental'' polarizabilities, and
   the dot-dashed line represents a calculation with the $O(Q^4)$
   polarizabilities. Experimental data from Ref. \protect{\cite{lucas}}
   also shown.}}}
\end{figure}

\begin{figure}[t,h,b,p]
   \vspace{0.5cm} \epsfysize=7.5cm
   \centerline{\epsffile{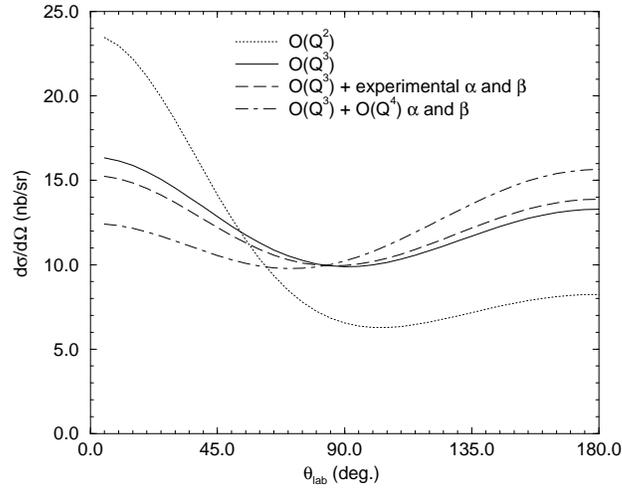}}
   \centerline{\parbox{11cm}{\caption{\label{fig-polcomp95} Results
   of calculations at 95 MeV using different values for the
   nucleon electromagnetic polarizabilities. Legend as in
   Figure~\protect{\ref{fig-polcomp49}}.}}}
\end{figure}

Two calculations were performed.  In the first, $\Delta \alpha$ and
$\Delta \beta$ were chosen so that the total polarizabilities
(\ref{mockpol}) were equal to the ``experimental'' values
(\ref{protpolexpt}), (\ref{neutpolexpt1}), and (\ref{neutpolexpt2}).
The second calculation involved a more dramatic change in the
polarizabilities: $\Delta \alpha$ and $\Delta \beta$ were chosen so
that $\alpha$ and $\beta$ were equal to the $O(Q^4)$ values of
Eqs.~(\ref{eq:alphaOQ4}) and (\ref{eq:betaOQ4}).  In either case
$\Delta \alpha_p+\Delta \alpha_n$ is relatively small, while $\Delta
\beta_p+\Delta \beta_n$ is not large for ``experimental'' values, but
is significant for the $O(Q^4)$ values.  The results of these two
calculations for the two photon energies $E_\gamma=49$ MeV and
$E_\gamma=95$ MeV are shown in Fig.~\ref{fig-polcomp49} and
Fig.~\ref{fig-polcomp95}.

\begin{figure}[t,h,b,p]
   \vspace{0.5cm} \epsfysize=7.5cm
   \centerline{\epsffile{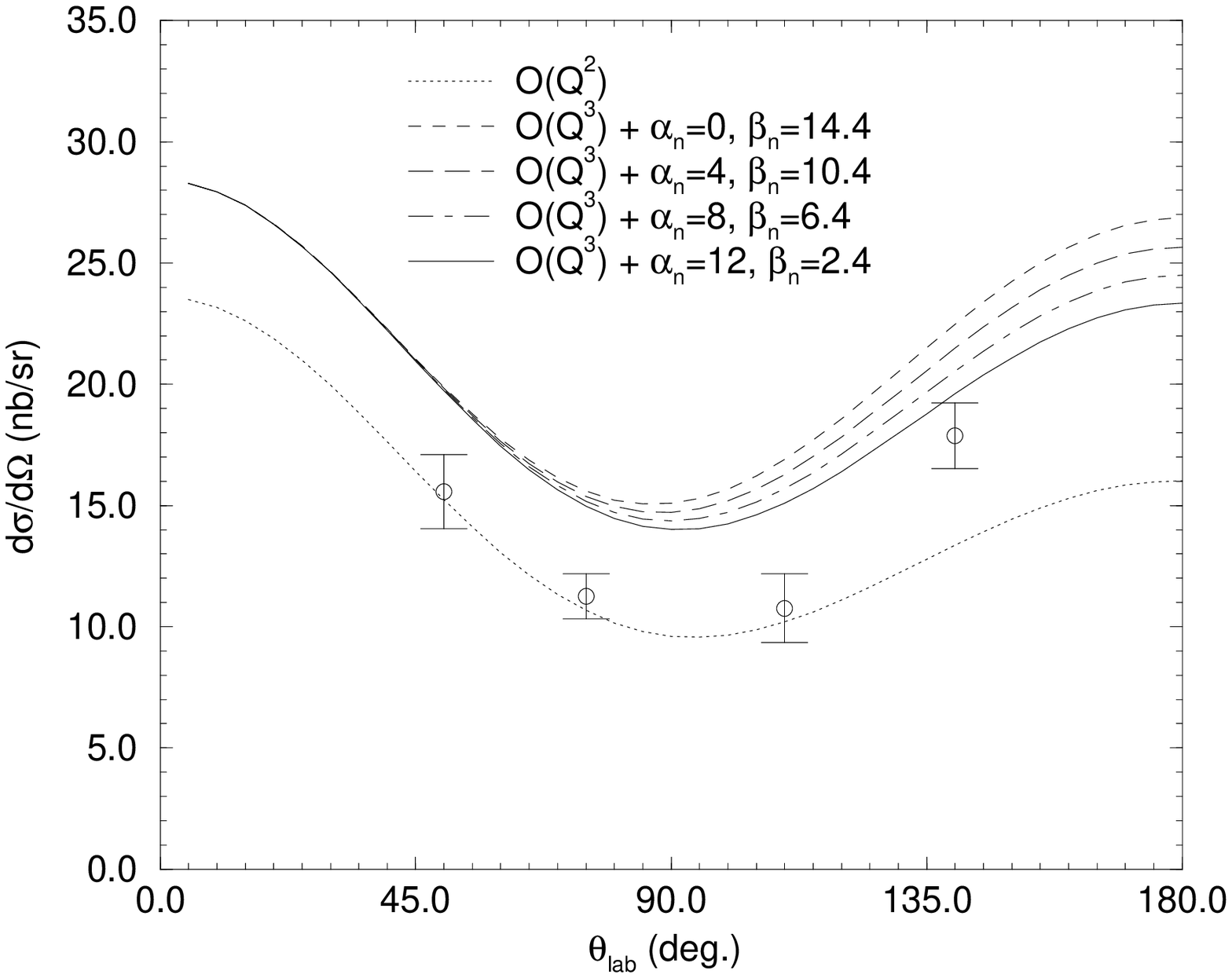}}
   \centerline{\parbox{11cm}{\caption{\label{fig-neutronfiddle49} 
   Results of calculations at 49 MeV which vary the neutron electric
   and magnetic polarizabilities while holding the sum fixed.}}}
\end{figure}

\begin{figure}[t,h,b,p]
   \vspace{0.5cm} \epsfysize=7.5cm
   \centerline{\epsffile{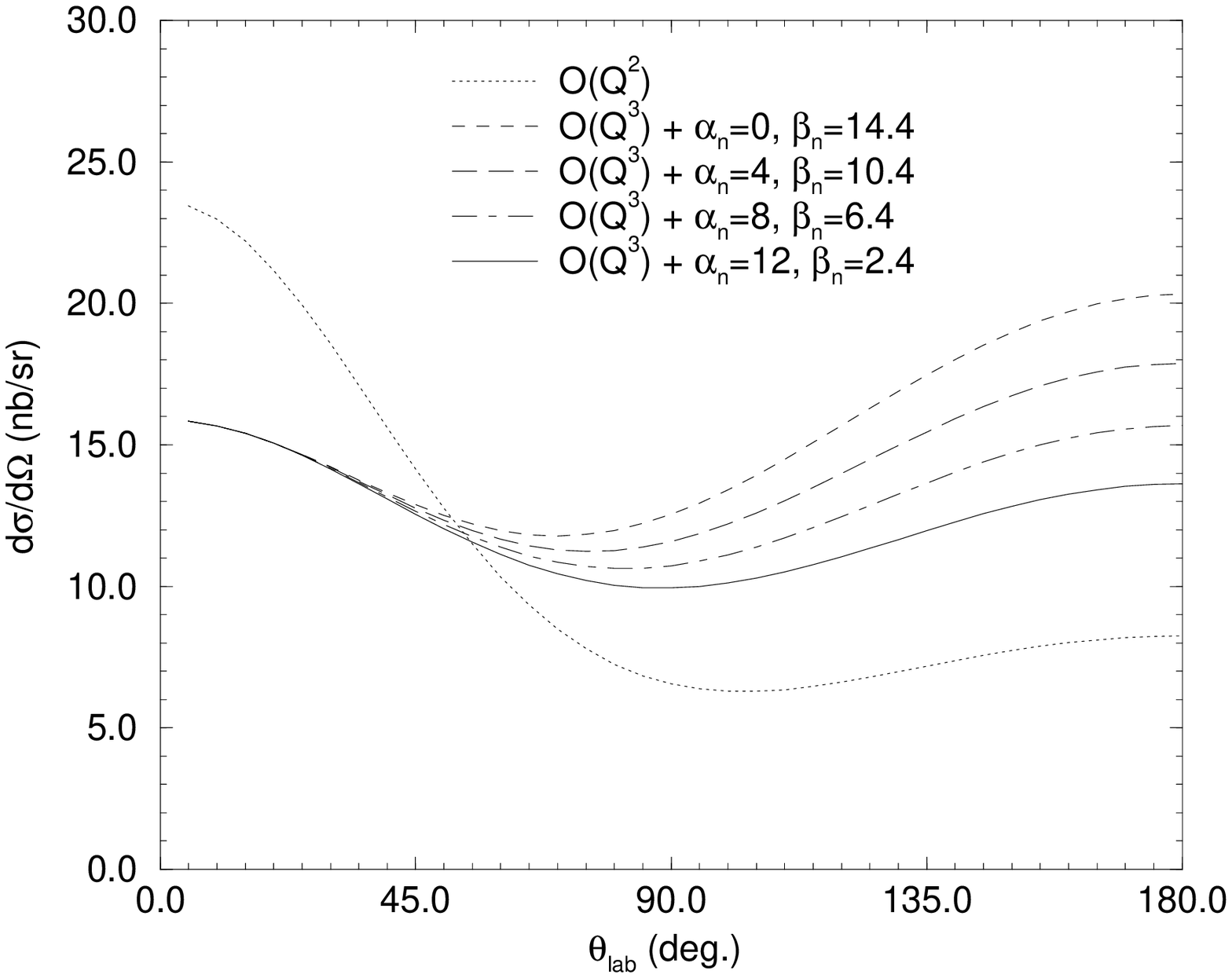}}
   \centerline{\parbox{11cm}{\caption{\label{fig-neutronfiddle95} 
   Results of calculations at 95 MeV which vary the neutron electric
   and magnetic polarizabilities while holding the sum fixed.}}}
\end{figure}

In both cases we see that, just as one would expect, the cross section
at 95 MeV is much more sensitive to these $O(Q^4)$ terms than the
cross section at 49 MeV.  It is not surprising that the calculation
with $O(Q^4)$ polarizabilities exhibits a larger change than that with
``experimental'' values.  Continually increasing $\beta_p+\beta_n$ at
approximately constant $\alpha_p+\alpha_n$ decreases the cross section
at forward angles and increases it at back angles.  In fact, it seems
that if $\beta_p+\beta_n$ is sufficiently large then the character of
the cross section at 95 MeV can change completely from forward peaked
to backward peaked. 

As a final exercise, Figs.~\ref{fig-neutronfiddle49} and
\ref{fig-neutronfiddle95} exhibit the differential cross sections
found at 49 and 95 MeV if $\alpha_p$ and $\beta_p$ are held fixed at
their $O(Q^3)$ values, while $\beta_n$ and $\alpha_n$ are varied with
$\alpha_n + \beta_n$ constrained to the central value of
Eq.~(\ref{neutpolexpt1}). These plots suggest that at backward angles
and higher energies the differential cross section is quite sensitive
to $\alpha_n - \beta_n$.

However, significant change in the 95 MeV results presented here from
those obtained at $O(Q^3)$ mandates a cautious interpretation.  At the
same time that the cross section at 95 MeV is more sensitive to
polarizabilities than at lower energies, it is also more sensitive to
$O(Q^4)$ corrections. In our view, a full $O(Q^4)$ calculation in
$\cpt$ is necessary if any attempt is to be made to extract the
neutron polarizability from the Saskatoon data within this framework.

\section{Conclusion}
\label{sec-conclusion}

We have calculated the differential cross section
for Compton scattering on the deuteron in $\cpt$ up to $O(Q^3)$.
We have found:

$\bullet$ Reasonable agreement with the data at 49 MeV.  At this
energy $O(Q^3)$ corrections are not large compared to the leading
$O(Q^2)$ result, and $O(Q^4)$ terms seem to be even smaller.  There is
little dependence on the wavefunction used.  However, as anticipated,
certain terms necessary to restore the Thomson, zero-energy, limit for
photon-deuteron scattering are important at this lower energy. Once
they are added to our calculation agreement with data is good.

$\bullet$ Good agreement with the data at 69 MeV.  At this energy the
convergence appears to be good. This suggests that $\cpt$ at $O(Q^3)$ is
providing reasonable neutron and two-nucleon contributions.

$\bullet$ That the polarizability approximation should not be used in
the calculation of the differential cross section at 95 MeV, since
truncating the photon-nucleon amplitude at order $\omega^2$ results in
a significant change in the photon-deuteron differential cross section
for forward angles.

$\bullet$ Wave function dependence on the order of 10\% in the
differential cross section. Our calculations suggest that much of this
dependence would be removed if the physics of pion range were the same
in both wave functions and in the computation of the kernel.

$\bullet$ A prediction at 95 MeV which is, however, plagued by
considerable uncertainties. Convergence is slow at this energy, as
indicated by the relative size of both the full set of $O(Q^3)$
corrections and a partial set of $O(Q^4)$ corrections.  The cross
section tends to come out somewhat smaller than at lower energies, in
particular in the backward directions, although the full $O(Q^4)$
amplitude is likely to be somewhat bigger at back angles.  It seems
that a more stringent test of $\cpt$ at these energies---including
aspects of neutron structure beyond the $O(Q^3)$ ``pion cloud''
picture---will have to wait for a next-order calculation.

\bigskip \bigskip

\section*{Acknowledgements}
Discussions with Jiunn-Wei Chen, Tom Cohen, Jim Friar, Harald
Grie{\ss}hammer, Dave Hornidge, Mark Lucas, Ulf Mei{\ss}ner, Martin
Savage, and Roxanne Springer are gratefully acknowledged.  We also
thank Vincent Stoks for providing us with the Nijm93 wave function.
This research was supported in part by the DOE grants
DE-FG02-93ER-40762 and DE-FG03-97ER41014 and by the National Science
Foundation grant PHY 94-20470. M.M would like to thank Capes and CNPq,
Brazil, for support.


\newpage



\newpage

\begin{thebibliography}{99}

\bibitem{bkmrev}  V.~Bernard, N.~Kaiser, and Ulf-G.~Mei{\ss}ner,
                  Int. J. Mod. Phys. {\bf E4}, 193 (1995), 
                  {\tt hep-ph/9501384}.

\bibitem{ulf1}    V.~Bernard, N.~Kaiser, and Ulf-G.~Mei{\ss}ner,
                  Phys. Rev. Lett. {\bf 67}, 1515 (1991);
                  Nucl. Phys. {\bf B383}, 442 (1992);
         V.~Bernard, N.~Kaiser, J.~Kambor, and Ulf-G.~Mei{\ss}ner,
               Nucl. Phys. {\bf B388}, 315 (1992).

\bibitem{ulf2}
V. Bernard, N. Kaiser, A. Schmidt, and Ulf-G. Mei{\ss}ner, 
            Phys. Lett. {\bf B319}, 315 (1993), {\tt hep-ph/9309211};
          Z. Phys. {\bf A348}, 317 (1994).

\bibitem{newanalysis} 
J. Tonnison, A.~M. Sandorfi, S. Hoblit, and A.~M. Nathan,
Phys. Rev. Lett. {\bf 80}, 4382 (1998), {\tt nucl-th/9801008}.

\bibitem{baldin}
A.~M. Baldin, Nucl. Phys. {\bf 18}, 310 (1960);
M. Damashek and F. Gilman, Phys. Rev. {\bf D1}, 1319 (1970);
D. Babusci, G. Giordano, and G. Matone, Phys. Rev. {\bf C57}, 291 (1998),
{\tt nucl-th/9710017}.

\bibitem{neutpol1} J. Schmiedmayer et al, 
                   Phys. Rev. Lett. {\bf 66}, 1015 (1991).

\bibitem{jerry} J.~J. Karakowski and G.~A. Miller, {\tt
    nucl-th/9901018}; J.~J. Karakowski, Ph.~D. thesis, University of
  Washington (1999), {\tt nucl-th/9901011}.

\bibitem{neutpol2} L. Koester et al, Phys. Rev. {\bf C51}, 3363 (1995).
 
\bibitem{neutpol3} K.~W. Rose et al, Nucl. Phys. {\bf A514}, 621 (1990).

\bibitem{wissmann} F. Wissmann, M.~I. Levchuk, and M. Schumacher,
Eur. Phys. J. {\bf A1}, 193 (1998).

\bibitem{hornidge} D. Drechsel, {\it et al.}, in {\it Mainz 1997, Chiral
Dynamics: Theory and Experiment}, ed. A.~Bernstein {\it et al}, 
(Springer-Verlag, 1998), p. 264, 
{\tt nucl-th/9712013}.

\bibitem{lucas}
M.~Lucas, Ph.~D. thesis, University of Illinois, unpublished (1994). 

\bibitem{SAL} D. Hornidge, private communication;
G.~Feldman, private communication.

\bibitem{lund} M.~Lunding, private communication.

\bibitem{wilbois} T. Wilbois, P. Wilhelm, and H. Arenh\"ovel,
                  Few-Body Systems Suppl. {\bf 9}, 263 (1995).

\bibitem{levchuk} M.~I. Levchuk and A.~I. L'vov, {\tt nucl-th/9809034},
and references therein.

\bibitem{weinnp}  S. Weinberg, Phys. Lett. {\bf B251}, 288 (1990); 
              Nucl. Phys. {\bf B363}, 3 (1991);
              Phys. Lett. {\bf B295}, 114 (1992).

\bibitem{ordonez} 
C. Ord\'{o}\~{n}ez and U. van Kolck, Phys. Lett. {\bf B291}, 459 (1992);
U. van Kolck, Ph.~D. thesis, University of Texas (1993).

\bibitem{vk} 
 U. van Kolck, Phys. Rev. {\bf C49}, 2932 (1994).

\bibitem{ray} 
 C. Ord\'{o}\~{n}ez, L. Ray, and U. van Kolck, 
                     Phys. Rev. Lett. {\bf 72}, 1982 (1994);
                     Phys. Rev. {\bf C53}, 2086 (1996), {\tt hep-ph/9511380}.

\bibitem{lepage} G.~P. Lepage, lectures given at the 
9th Jorge Andre Swieca Summer School: Particles and Fields, Sao Paulo,
Brazil, {\tt nucl-th/9706029}.

\bibitem{seki} 
{\it Nuclear Physics with Effective Field Theory},
ed. R. Seki, U. van Kolck, and M.~J. Savage,
                 World Scientific (1998).

\bibitem{pc} 
U. van Kolck, in 
{\em Mainz 1997, Chiral Dynamics: Theory and 
Experiment\/}, ed. A. Bernstein {\it et al}, (Springer-Verlag,1998),
{\tt hep-ph/9711222}; in Ref. \cite{seki}; 
Nucl. Phys. {\bf A645}, 273 (1999), {\tt nucl-th/9808007}.

\bibitem{ksw}
D.~B. Kaplan, M.~J. Savage, and M.~B. Wise, Phys. Lett. {\bf B424}, 390 (1998);
{\tt nucl-th/9801034}; 
Nucl. Phys. {\bf B534}, 329 (1998), {\tt nucl-th/9802075}.

\bibitem{martinetal}
 J.-W.~Chen, H.~W.~Griesshammer, M.~J.~Savage and R.~P.~Springer,
{\tt nucl-th/9806080}; {\tt nucl-th/9809023};  
 J.-W.~Chen, {\tt nucl-th/9810021}.  

\bibitem{monster} U. van Kolck, {\tt nucl-th/9902015}.

\bibitem{silas1} S.~R.~Beane, V.~Bernard, T.-S.~H.~Lee and Ulf-G.~Mei{\ss}ner,
 Phys. Rev. {\bf C57}, 424 (1998), {\tt nucl-th/9708035}.

\bibitem{silas2} S.~R.~Beane, C.~Y.~Lee, and U.~van Kolck,
                 Phys. Rev. {\bf C52}, 2914 (1995), {\tt
                 nucl-th/9506017}; S.R.~Beane, V.~Bernard, T.S.H.~Lee,
                 Ulf-G.~Mei{\ss}ner and U.~van Kolck, Nucl. Phys. {\bf
                 A618}, 381 (1997), {\tt hep-ph/9702226}.

\bibitem{rho1} T.-S. Park, K. Kubodera, D-P. Min, and M. Rho,
Phys. Rev. {\bf C58}, 637 (1998), {\tt hep-ph/9711463}.

\bibitem{rho2} T.-S. Park, K. Kubodera, D-P. Min, and M. Rho, {\tt
astro-ph/9804144}; Nucl. Phys {\bf A646}, 83 (1999), {\tt
nucl-th/9807054}.

\bibitem{Friar} J.~L. Friar, Few-body Systems {\bf 22}, 161 (1997),
{\tt nucl-th/9607020}.

\bibitem{FT81} J.~L. Friar and E.~L. Tomusiak, Phys. Lett. {\bf B122}, 
11 (1983).

\bibitem{jenkinsetal} E.~Jenkins and A.~V.~Manohar,
  Phys. Lett. {\bf B255}, 558 (1991).

\bibitem{hemmert} M.~N Butler and M.~J. Savage, Phys. Lett. {\bf
                  B294}, 369 (1992); T.~R. Hemmert, B.~R. Holstein,
                  and J. Kambor, Phys. Rev. {\bf D55}, 5598 (1997),
                  {\tt hep-ph/9612374}; Phys. Rev. {\bf D57}, 5746
                  (1998), {\tt nucl-th/9709063}.

\bibitem{babusci} D. Babusci, G. Giordano, and G. Matone,
                  Phys. Rev. {\bf C55}, R1645 (1997).

\bibitem{KW} J.~H. Koch and R.~M. Woloshyn, 
                  Phys. Rev. {\bf C16}, 1968 (1977).

\bibitem{Bonnrep} R. Machleidt, K. Holinde, and Ch. Elster,
Phys. Rep. {\bf 149}, 1 (1987).

\bibitem{Nijm93} V.~G.~J. Stoks, R.~A.~M. Klomp, C.~P.~F. Terheggen,
and J.~J. de Swart, Phys. Rev. {\bf C49}, 2950 (1994), {\tt
nucl-th/9406039}.

\bibitem{willhelm} P. Wilhelm, Phys. Rev. {\bf C56}, 1215
(1997), {\tt nucl-th/9703037}.
\end{thebibliography}
\end{document}